\documentclass[conference]{IEEEtran}
\IEEEoverridecommandlockouts

\usepackage{cite}
\usepackage{amsmath,amssymb,amsfonts}
\usepackage{algorithm}
\usepackage{algorithmic}
\usepackage{graphicx}
\usepackage{textcomp}
\usepackage{xcolor}

\usepackage{booktabs}   
\usepackage{subcaption} 
\usepackage{pifont}
\usepackage{bm}

\def\BibTeX{{\rm B\kern-.05em{\sc i\kern-.025em b}\kern-.08em
    T\kern-.1667em\lower.7ex\hbox{E}\kern-.125emX}}
\begin{document}

\title{SplitFT: An Adaptive Federated Split Learning \\
System For LLMs Fine-Tuning}

\author{

\IEEEauthorblockN{Yimeng Shan}
\IEEEauthorblockA{
\textit{Hong Kong Polytechnic University}\\
Hong Kong \\
yi-meng.shan@connect.polyu.hk} 
\and
\IEEEauthorblockN{Zhaorui Zhang}
\IEEEauthorblockA{
\textit{Hong Kong Polytechnic University}\\
Hong Kong \\
zhaorui.zhang@polyu.edu.hk}
\and
\IEEEauthorblockN{Sheng Di}
\IEEEauthorblockA{
\textit{Argonne National Labratory}\\
USA \\
sdi1@anl.gov}
\and
\IEEEauthorblockN{Yu Liu}
\IEEEauthorblockA{
\textit{Hong Kong Polytechnic University}\\
Hong Kong \\
yu-y.liu@polyu.edu.hk}
\and
\IEEEauthorblockN{Xiaoyi Lu}
\IEEEauthorblockA{
\textit{University of California, Merced}\\
USA \\
xiaoyi.lu@ucmerced.edu}
\and
\IEEEauthorblockN{Benben Liu}
\IEEEauthorblockA{
\textit{The University of Hong Kong}\\
Hong Kong \\
benbenliu@hku.hk}




}

\maketitle

\begin{abstract}
Federated Split Learning has been identified as an efficient approach to address the computation resource constraints issue of the clients for classical federated learning, while guaranteeing data privacy for distributed model training across data owners. However, it faces some critical challenges when such a training strategy meets large language models (LLMs) for fine-tuning. Such challenges include setting the cutlayer adaptively across different clients to address the data and device heterogeneity issues, which affect the system performance significantly. In addition, efficiently reducing the communication overhead during the fine-tuning procedure is also another challenge. No work tries to address these challenges. 

To bridge this gap, we propose \textit{SplitTF}, an adaptive federated split learning system for LLMs fine-tuning. \textit{SplitFT} enables different clients to set different cut layers according to their computation resources and trained model performance. \textit{SplitFT} also proposes to reduce the LoRA rank in cutlayer to reduce the communication overhead. In addition to simulating the heterogeneous data in real-world applications for our proposed split federated learning system, we propose a length-based Dirichlet approach to divide the training data into different clients. Extensive experimental results show that our proposed approach outperforms the state-of-the-art approach for fine-tuning time efficiency and model performance based on various popular benchmarks.
\end{abstract}


\section{Introduction}

In recent years, large-scale models have achieved significant advancements across various domains, including computer vision, natural language processing, and life sciences \cite{li2024scopingreviewusinglarge,wangdan,ma2025moe,ma2025compression}, with particular emphasis on large language models (LLMs) \cite{openai2024gpt4technicalreport}. The field of generative AI, especially LLMs, has experienced remarkable growth, catalyzing the rapid development of numerous innovative applications such as LLM chatbots, programming assistants, image and video generation, and writing aids \cite{c:1}. Researchers and institutions have contributed to this progress by making their pre-trained models publicly accessible on platforms like Hugging Face \cite{vaswani2023attentionneed}. These models are predominantly trained on publicly available data sourced from the Internet, enabling them to produce high-quality results and excel in non-domain-specific applications \cite{c:2, informatics11030057}.

Pre-trained models encounter several challenges. Firstly, while these models demonstrate high accuracy in non-domain-specific applications, they often exhibit reduced accuracy when applied to domain-specific fields such as medicine and law. Secondly, the quality of training data is a critical factor that constrains model performance. Recent studies indicate that the availability of high-quality public data is diminishing. Consequently, private data is garnering increasing attention from researchers and companies as a means to enhance model quality. The proliferation of open-source pre-trained LLMs presents an opportunity to fine-tune these models for specific domains that involve private data, thereby improving their capabilities in specialized areas. As a result, fine-tuning pre-trained models for vertical domains has become a prominent research focus. To mitigate computational overhead for the large-scale model training, several parameter-efficient fine-tuning (PEFT) \cite{peft, liu2025hlora,zhang2025cllora} strategies have been developed, including prompt learning \cite{Lester2021}, adapters \cite{Houlsby2019}, and low-rank adaptation (LoRA) \cite{hu2022lora}. These strategies involve freezing the pre-trained parameters of LLMs and training a small set of additional parameters. LoRA, in particular, has gained widespread adoption across various applications.
Federated learning offers a promising approach for fine-tuning LLMs across disparate data silos containing private data \cite{c:9}, \cite{xu2024fedfa, zhang2025fedcspc,zhang2025fedefsz}. This method facilitates the exchange of model parameters rather than the actual training data, thereby protecting data privacy \cite{c:3}. To fine-tune pre-trained LLMs in specific domains using private data, it is necessary to aggregate data from various data silos, where direct data transfer is not feasible. However, deploying federated learning across a vast number of devices presents challenges, particularly due to the limited computational and communication resources available \cite{zhang2022mipd, zhang2021sapus, zhaorui2022momentum}. This is especially pertinent for fine-tuning LLMs, which is inherently a resource-intensive process.


Split learning \cite{vepakomma2018splitlearninghealthdistributed} presents a viable solution to the challenges encountered in cross-device federated learning. This approach enables clients to train only a minimal number of model layers, while the server handles the training of the larger portion of the model. This division significantly reduces the computational burden on the client while also maintaining data privacy. Split learning partitions a comprehensive machine learning model into smaller network segments, allowing clients to focus on training specific modules of large language models (LLMs). This approach markedly decreases the processing demands on devices with limited resources, as opposed to executing the entire model \cite{c:6}. Furthermore, split learning enhances privacy by ensuring that clients do not have access to the server-side model and vice versa, thus protecting sensitive information throughout the training process \cite{c:7}.

Implementing split learning within federated learning to fine-tune LLMs across various data silos or devices that handle sensitive data presents several challenges. First is device heterogeneity, where the computational resources available across different devices vary significantly. During each round of fine-tuning, high-performance devices must wait for slower ones, which can severely impact the overall fine-tuning efficiency. The second challenge is data heterogeneity, where the data held by each device differs substantially. This variation can significantly affect the model's performance, as the diverse data distributions may lead to inconsistencies in the fine-tuning outcomes. Thirdly, during each fine-tuning round, the client is required to transmit the intermediate "smashed" data to the server, which introduces substantial communication overhead and can adversely affect the overall system performance.

To address the aforementioned challenges, we design and implement \textit{SplitFT}, an efficient and adaptive federated split learning system for fine-tuning LLMs utilizing the LoRA parameter-efficient fine-tuning strategy. \textit{SplitFT} facilitates the fine-tuning of LLMs on devices with limited resources within federated learning environments. To enhance fine-tuning performance, \textit{SplitFT} incorporates a heuristic algorithm that adaptively adjusts layer allocation based on the computational resources available to each client. Additionally, \textit{SplitFT} proposes reducing the LoRA rank for the cutlayer to minimize communication overhead, thereby improving system performance. To further assess the impact of data heterogeneity, \textit{SplitFT} provides a length-based Dirichlet approach to partition the fine-tuning dataset. \textit{\textbf{Our contributions are outlined as follows:}}

\begin{itemize}
    \item We design and implement \textit{SplitFT}, a robust system tailored for fine-tuning LLMs within federated learning environments. \textit{SplitFT} enables the fine-tuning of LLMs on devices with limited resources, while protecting the data privacy of each device and ensuring the performance and quality of the model. \textit{SplitFT} is designed to be modular and extensible, facilitating the integration of various algorithms, data types, sampling techniques, and model configurations. 
    
    \item We make a unique contribution by designing a heuristic algorithm that adaptively allocates layers based on the data and resource heterogeneity of each device within a federated learning environment. This strategy ensures an optimal balance between the data processing tasks on clients and the computational load on the server after each global round. In contrast to baseline approaches that utilize fixed layer splits with independent and identically distributed (IID) data, our method achieves superior model performance and faster convergence, particularly in scenarios characterized by heterogeneous data distributions. To further reduce communication overhead, we assign a smaller LoRA rank to the cutlayer, while employing a larger LoRA rank for other layers to maintain high quality and performance of the models.

    \item We provide a length-based Dirichlet approach specifically designed for the context of dataset partitioning in LLMs training and fine-tuning scenarios. This method aims to further assess the impact of data heterogeneity on model performance within federated learning environments. The dataset is divided into $N$ classes based on data length, and these classes are distributed among clients using the Dirichlet approach and various according to the hyperparameter $\alpha$.

    \item We conducted a comprehensive evaluation of our proposed \textit{SplitFT} using several popular LLM benchmarks. The experimental results indicate that \textit{SplitFT} outperforms the current state-of-the-art approaches across various benchmarks, underscoring its efficiency and effectiveness.

\end{itemize}


\section{Background and Motivations}
\label{background}
\subsection{Federated Split Learning}

Federated Split Learning effectively capitalizes on the benefits of Federated Learning by facilitating parallel training across multiple clients, while simultaneously reducing communication overhead and addressing limited computational resources through model partitioning. This approach differentiates itself from traditional Federated Learning and Split Learning by integrating their strengths to optimize model training in resource-constrained environments. We employ Federated Split Learning in conjunction with FedAvg \cite{li2019convergence}. The workflow of Federated Split Learning is outlined as follows.


The Federated Split Learning process begins with the central server activating a designated number \( N \) of clients to participate in the Split Federated Learning process. The complete machine learning model is denoted as \( W \), which is partitioned into two segments: the server-side model \( W^s \) and the client-side model \( W^c \), satisfying the relationship as $W = W^s + W^c$. 



Here, \( W^s \) represents the portion of the model residing on the server, while \( W^c \) is maintained by each client. During a local training round, each client performs forward propagation using its local model segment \( W^c \), generating smashed data that is sent to the server. The server then computes the parameter update gradient for \( W^s \) based on its loss function and returns the gradient of the smashed data to the client. Utilizing this gradient, the client computes its parameter update gradient to adjust \( W^c \). This process constitutes a local round of training. Following the local training rounds, the updated client model segments \( W^c \) are sent back to the FedServer, which aggregates these updates \( \delta \) using the FedAvg algorithm. FedAvg weighs each client's update based on the number of samples they possess, ensuring a balanced and representative aggregation of the global model. The aggregated updates are then used to update the server’s base model \( W \). The accuracy of the aggregated model \( W \) is evaluated during the testing phase to assess its performance. Subsequently, the updated global model \( W \) is redistributed to the clients for further refinement, continuing the iterative and privacy-preserving training cycle.

The integration of FedAvg \cite{li2019convergence} within Split Federated Learning provides several notable advantages. \textit{Firstly}, by partitioning the model, Split Federated Learning ensures that clients manage only a portion of the model, thereby reducing the risk of privacy leakage compared to traditional federated learning. \textit{Secondly}, this model partitioning limits the volume of data exchanged between clients and the server, enhancing communication efficiency during the training process. \textit{Lastly}, Split Federated Learning accommodates the limited computational resources of each device by enabling clients to train smaller model segments, thereby optimizing resource utilization. Through this integrated approach, Split Federated Learning with FedAvg offers a robust solution for training large language models (LLMs) on resource-constrained devices, effectively balancing model performance with resource and communication efficiency. This framework addresses the inherent limitations of both federated learning and split learning, providing enhanced privacy, reduced communication overhead, and optimized resource utilization, making it particularly well-suited for deployment in heterogeneous and resource-limited environments.


\subsection{Parameter-Efficient Fine-Tuning (PEFT)}

Traditional full-model fine-tuning and training pose significant challenges for edge and resource-constrained Internet of Things (IoT) devices due to the substantial demands for parameter storage, intensive computation, and frequent communication. These challenges arise from the requirement to store and compute gradients for all model parameters. While Split Federated Learning improves model convergence through parallelized communication and model partitioning, full-parameter training continues to impose considerable communication overhead, thereby hindering training efficiency. Consequently, this approach restricts the model's ability to rapidly achieve optimal performance in downstream tasks.

To address these challenges, parameter-efficient fine-tuning techniques have been developed. However, widely adopted methods such as prefix tuning and adapter tuning remain resource-intensive, imposing substantial storage and computational burdens that limit their practical applicability in resource-constrained environments. Adapter tuning introduces additional adapter layers with dimensionality-reduction and dimensionality-expansion matrices, while prefix tuning incorporates prefix vectors independent of the model's original weight matrices. Both methods significantly increase computational overhead and contribute to a substantial increase in the network's effective depth. In contrast, Low-Rank Adaptation (LoRA) circumvents the insertion of additional layers into the original model structure by applying low-rank approximations directly to the weights within existing layers. This approach reduces the additional computational complexity, particularly in deep large language models (LLMs). Consequently, in this work, we adopt the LoRA method to effectively address computational resource constraints.

Pretrained models' weight matrices \( W_{0} \in \mathbb{R}^{d \times k} \) are known to have a low intrinsic dimensionality. The LoRA method leverages this property by decomposing the trainable matrices according to their low intrinsic rank as follows formula (\ref{lora}):

\begin{equation}
W = W_0 + \Delta W = W_0 + AB
\label{lora}
\end{equation}

Where \( A \in \mathbb{R}^{d \times r} \) and \( B \in \mathbb{R}^{r \times k} \) with \( r \ll \min(d, k) \). This method not only limits the number of trainable parameters to approximately 1\% by the amount of \( \Delta W = d \times r + r \times k \), but also significantly reduces the gradient size during backpropagation. In federated split learning, frequent communication is necessary, and the LoRA method alleviates this burden by reducing the size of gradients exchanged between clients and the server. We integrate LoRA into our \textit{SplitFT} due to its computational efficiency and its suitability for resource-constrained environments.

\section{The Design of \textit{SplitFT}}
\label{system_design}
\subsection{Main Components of \textit{SplitFT}}

\textbf{\textit{LoRA Adapter.}} In \textit{SplitFT}, both the server- and client-side pretrained models are structured similarly to the GPT-series models. Each layer comprises two primary components: \textit{GPTAttn} and \textit{GPTMLP}, which together form a \textit{GPTBlock}. To reduce the number of trainable parameters, we apply \textit{Low-Rank Adaptation (LoRA)} to each layer. Within the context of split learning, the \textbf{cutlayer} serves as the communication boundary between the client and server. The cutlayer is defined as the last layer on the client-side and the first layer on the server-side. The LoRA rank for the cutlayer is denoted as \( r_{\text{cut}} \), while the LoRA ranks for all other layers on both the client-side and server-side are denoted as \( r_{\text{others}} \). By adjusting these ranks, we can manage the trade-off between model performance and communication efficiency, ensuring that critical layers maintain high adaptability while non-critical layers remain efficient.


\textbf{\textit{Client.}} Each client in the \textit{SplitFT} is assumed to possess sufficient computational resources and adequate storage capacity for handling both parameters and gradients during forward and backward propagation. The dataset for each client \( i \) is represented as \( \mathcal{D}_i = \left\{ (\mathbf{x}_{i,j}, y_{i,j}) \right\}_{j=1}^{D_i} \), where \( \mathbf{x}_{i,j} \) and \( y_{i,j} \) denote the \( j \)-th batch of input data and its corresponding labels for client \( i \), respectively. The weight matrix of the client-side pretrained models is denoted as \( W_c \). We assume that the client-side pretrained model consists of \( m \) layers, whereas the entire pretrained LLM model comprises \( M \) layers. The client-side LoRA adapters for client \( i \) are denoted as: \(O_{c,i} = \left\{ \mathbf{A}_{c,i}^{1}, \mathbf{B}_{c,i}^{1}, \mathbf{A}_{c,i}^{2}, \mathbf{B}_{c,i}^{2}, \dots, \mathbf{A}_{c,i}^{m}, \mathbf{B}_{c,i}^{m} \right\}\), where \( \mathbf{A}_{c,i}^{p} \) and \( \mathbf{B}_{c,i}^{p} \) represent the decomposition matrices of the \( p \)-th LoRA adapter for client \( i \), and \( p \) ranges from 1 to \( m \). It is important to highlight that the LoRA rank for the cutlayer (\( r_{\text{cut}} \)) is applied to the last layer of the client-side model, while all other layers use a rank of \( r_{\text{others}} \). This configuration ensures that the most critical layer for communication maintains high adaptability, balancing performance and communication efficiency.

\textbf{\textit{Main Server.}} The main server in \textit{SplitFT} is equipped with sufficient computational power to manage the remaining layers of the pretrained model, enabling LoRA fine-tuning and conducting forward and backward propagation for these layers effectively. The server-side pretrained model consists of two components: the server model and the base model. The server model comprises the remaining \( M - m \) layers of the entire LLM, excluding the \( m \) layers handled by the client-side pretrained models. The base model includes the full pretrained model with \( M \) layers and is represented by the weight matrix \( W_{base} \), which can be expressed as: \(W_{base} = [\mathbf{W}_s ; \mathbf{W}_c]\), where \( \mathbf{W}_s \) represents the server-side model weights and \( \mathbf{W}_c \) represents the client-side model weights. Based on this partitioning, the set of server-side LoRA adapters is defined as: \(O_s = \left\{ \mathbf{A}_s^{1}, \mathbf{B}_s^{1}, \mathbf{A}_s^{2}, \mathbf{B}_s^{2}, \dots, \mathbf{A}_s^{M-m}, \mathbf{B}_s^{M-m} \right\}\), where \( \mathbf{A}_s^{p} \) and \( \mathbf{B}_s^{p} \) represent the decomposition matrices of the \( p \)-th LoRA adapter on the server side. Notably, \( \mathbf{A}_s^{1} \) and \( \mathbf{B}_s^{1} \) correspond to the LoRA adapters at the cutlayer and are assigned a rank of \( r_{\text{cut}} \). All other adapters on the server side utilize a rank of \( r_{\text{others}} \), ensuring a balance between model performance and communication efficiency.

\textit{\textbf{Local FedAvg Server.}} The Local FedAvg Server plays a critical role in the \textit{SplitFT} by receiving updates from each client-side LoRA adapter. These updates are aggregated based on the proportion of local data each client holds relative to the entire dataset. The aggregated updates are then sent to client and the main server to update the clients' server-side models and the corresponding LoRA adapters within the client-side pretrained models' layers.

In federated split learning, configuring the fully connected layers on the client side with a sufficiently large number of nodes substantially mitigates the risk of inferring client-side model parameters and raw data. This design choice enhances privacy by ensuring that the information exchanged during the training process does not reveal sensitive client data or model parameters. Furthermore, the aggregation process employs the FedAvg algorithm to maintain a balanced and representative global model, thereby enhancing the robustness and privacy-preserving capabilities of the \textit{SplitFT}.

\begin{figure*}
\centering
\includegraphics[width=0.9\linewidth, height=0.66\linewidth]{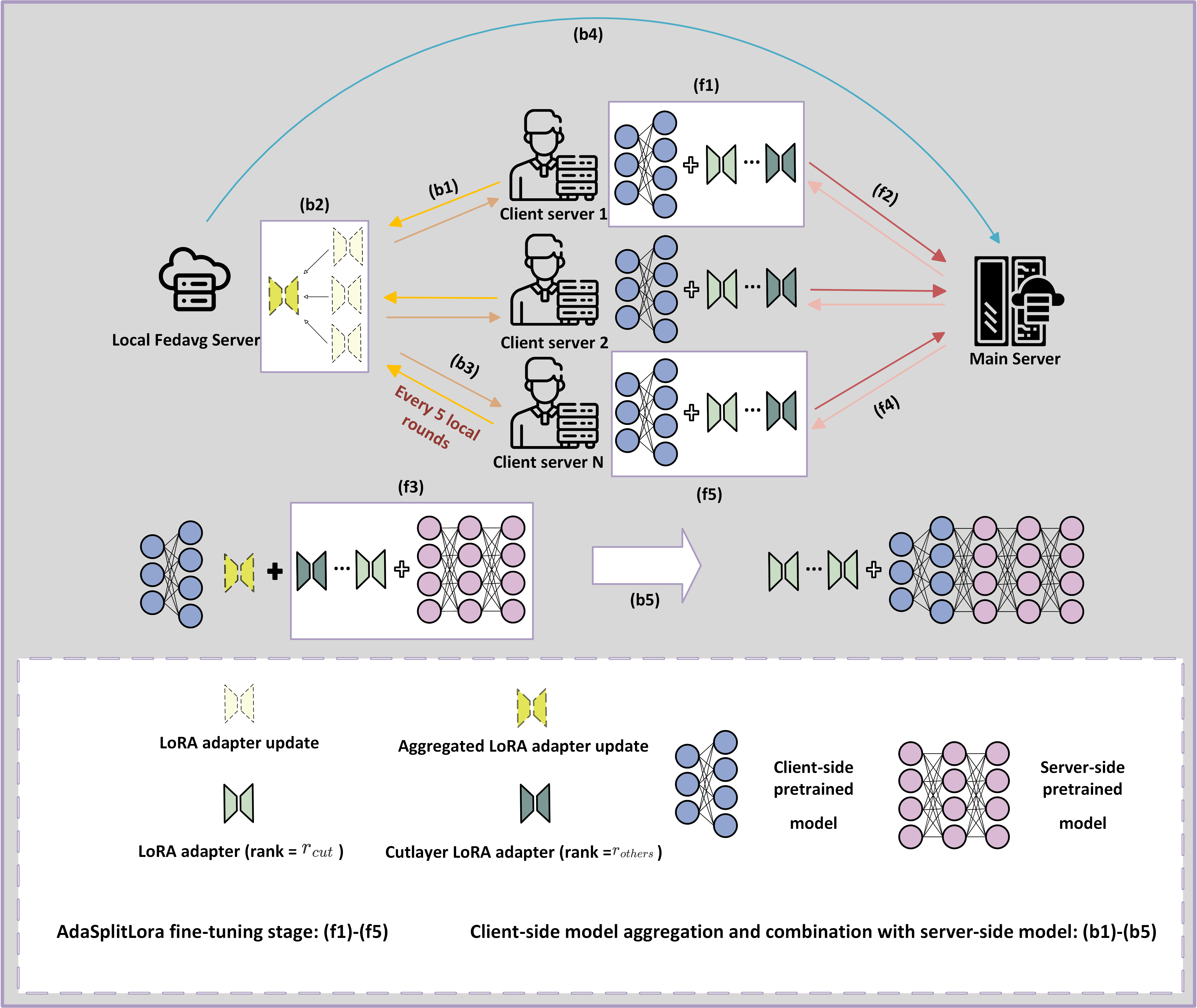} 
\caption{The System Overview of Our Proposed \textit{SplitFT}.}
\label{fig1}
\end{figure*}


\begin{algorithm}[!ht]
\caption{The Workflow for \textit{SplitFT} Fine-Tuning}
\label{alg:SplitFT}
\textbf{Input}: Initialized client-side models \( O_{c,i} \), server-side model \( O_s \), global base model \( W_{base} \), learning rates \( \gamma_c, \gamma_s \), client list \( N_{\text{client}} \), dataset \( \mathcal{D} \), Dirichlet parameter \( \alpha \), total global rounds \( R \)\\
\textbf{Parameter}: Number of client layers \( l_{c,i} \), server layers \( M-m \), adjustment weight \( \gamma \)\\

\noindent \textbf{Output}: Optimized global model \( W_{base} \)

\begin{algorithmic}[ht!]
\STATE \textbf{Initialization:} 
\STATE Partition \( \mathcal{D} \) using IID/Non-IID strategy (Dirichlet parameter \( \alpha \)).
\STATE Activate \( N_{\text{client}} = \{1, 2, \dots, N\} \) and initialize models \( O_{c,i}, O_s, W_{base} \).
    
\FOR{global round \( r = 1, \dots, R \)}
    \STATE \textbf{Client-side Fine-tuning:}
    \FOR{each client \( i \in N_{\text{client}} \)}
        \STATE Select a mini-batch \( D_{c,i}^r \) from \( \mathcal{D}_i \).
        \STATE Perform forward propagation to compute smashed data \( \mathbf{\varphi}_i^r \).
        \STATE Transmit \( \mathbf{\varphi}_i^r \) to the server.
    \ENDFOR

    \STATE \textbf{Server-side Training:}
    \STATE Compute predictions \( y_{pre}^r \) using \( \mathbf{\varphi}_i^r \).
    \STATE Compute server-side gradients \( \mathbf{g}_{A,s}, \mathbf{g}_{B,s} \) and update \( O_s \) with \( \gamma_s \).
    \STATE Transmit gradients \( \mathbf{g}_{\phi^r} \) back to clients.

    \STATE \textbf{Client-side Backward Propagation:}
    \FOR{each client \( i \in N_{\text{client}} \)}
        \STATE Compute gradients \( \mathbf{g}_{A,c,i}, \mathbf{g}_{B,c,i} \) using \( \mathbf{g}_{\phi^r} \).
        \STATE Update client-side LoRA adapters \( O_{c,i} \) with \( \gamma_c \).
    \ENDFOR

    \STATE \textbf{Client-side LoRA Adapter Aggregation:}
    \STATE Transmit updates \( \Delta O_c \) to the aggregation server.
    \STATE Perform weighted aggregation of LoRA updates.
    
    \STATE \textbf{Layer Adjustment and Model Update:}
    \STATE Adjust \( l_{c,i} \) for each client based on test accuracy \( \text{acc}_i \) and average accuracy \( \text{acc}_{\text{avg}} \).
    \STATE Update \( W_{base} \) and allocate new layers to clients.
\ENDFOR

\STATE \textbf{return} Optimized global model \( W_{base} \)
\end{algorithmic}
\end{algorithm}

\subsection{Dirichlet Strategies for Dataset Partitioning}
To evaluate the robustness of \textit{SplitFT}, we employed both Independent and Identically Distributed (IID) and Non-IID strategies to partition the text dataset. The IID strategy involves randomly sampling the entire dataset and dividing it into \( N \) equal portions, where \( N \) corresponds to the number of clients. Specifically, the set of clients is defined as \( N_{\text{client}} = \{1, 2, \ldots, N\} \), and the IID method partitions the dataset into \( N \) local client datasets accordingly.

For the Non-IID strategy, we propose a length-based Dirichlet partitioning approach. First, the original text dataset is tokenized, converting it into token streams or word/sentence representations. These tokens are then categorized into \( K \) predefined categories based on criteria such as topics, word frequency distributions, or semantic clusters, resulting in \( \mathcal{D} = \{D_1, D_2, \ldots, D_K\} \), where each \( D_k \) represents the set of samples belonging to category \( k \). For each category \( k \), a length-\( N \) vector \( \mathbf{p}_k = [p_{k1}, p_{k2}, \ldots, p_{kN}] \) is sampled from a Dirichlet distribution \( \mathbf{p}_k \sim \text{Dirichlet}(\boldsymbol{\alpha}) \), where \( \boldsymbol{\alpha} = [\alpha, \alpha, \ldots, \alpha] \) is the concentration parameter vector. Here, \( p_{ki} \) represents the proportion of category \( k \) assigned to client \( i \), subject to the constraints \( \sum_{i=1}^{N} p_{ki} = 1 \) and \( p_{ki} \geq 0 \). The total number of samples in category \( k \) is \( n_k = |D_k| \), and the number of samples allocated to client \( i \) from category \( k \) is \( n_{ki} = \lfloor p_{ki} \cdot n_k \rfloor \). Using these values, samples from \( D_k \) are randomly selected and allocated to the respective clients. Each client's local dataset is then formed by combining the allocated samples from all categories: \( D_{c,i} = \bigcup_{k=1}^{K} D_{ki} \), where \( D_{ki} \) represents the subset of samples from category \( k \) assigned to client \( i \).

The concentration parameter \( \alpha \) plays a significant role in determining the degree of heterogeneity in the Non-IID data distribution. A smaller \( \alpha \) (e.g., \( \alpha < 1 \)) results in a more skewed distribution, where each client predominantly receives data from only a few categories, creating high data heterogeneity. Conversely, a larger \( \alpha \) (e.g., \( \alpha > 1 \)) leads to a more uniform distribution, where data from each category is distributed more evenly across clients. As \( \alpha \) approaches infinity, the distribution converges to the IID scenario, where all clients receive nearly equal proportions of all categories.

\subsection{Adaptive Layer Allocation Strategy}
Due to the heterogeneity of the computation resources across different clients. If we adopt the FedAvg strategy to aggregate the training results from all clients, the faster clients will need to wait for the slower ones, which will cause performance degradation. Therefore, \textit{SplitFT} proposes to allocate a different number of layers for different clients according to their resources and model quality to guarantee the overall fine-tuning performance. The optimization objective for adaptive layer allocation for different clients can be formulated as the following formula (\ref{adap_layer}).

\begin{equation}
    \min_{\mathbf{O}_{c,i}, \mathbf{O}_s} \quad \sum_{i=1}^N w_i \cdot \frac{\lvert \mathcal{D}_{c,i} \rvert}{\lvert \mathcal{D} \rvert} L_i(W_{base} \mid \mathbf{O}_{c,i}, \mathbf{O}_s)
    \label{adap_layer}
\end{equation}

Where \( w_i \) is a dynamic adjustment weight factor that controls the training load of each client based on their performance. The weights are adjusted according to the following rules:

\noindent \textbf{\textit{Rules.}} Define \( \text{acc}_i \) as the test accuracy of client \( i \) during the current global round and \( \text{acc}_{\text{avg}} \) as the average test accuracy of all clients. 

\begin{itemize}
    \item If \( \text{acc}_i > \text{acc}_{\text{avg}} \), increase \( w_i \) as follows: \(w_i = 1 + \gamma (\text{acc}_i - \text{acc}_{\text{avg}})\), where \( \gamma \) is a control factor regulating the degree of weight adjustment. 
    \item If \( \text{acc}_i < \text{acc}_{\text{avg}} \), decrease \( w_i \) as follows: \(w_i = 1 - \gamma (\text{acc}_{\text{avg}} - \text{acc}_i).\), where \( \gamma \) is a control factor regulating the degree of weight adjustment. 
\end{itemize}

This dynamic layer adjustment mechanism ensures that clients possessing higher-quality data and superior performance assume greater computational responsibilities, thereby enhancing the overall model performance. Conversely, clients with lower-quality data experience a reduction in computational load, thereby minimizing their impact on the global model.

The detailed workflow of our proposed \textit{SplitFT} is shown in the above Fig. \ref{fig1} and the algorithm \ref{alg:SplitFT}. In a local training round, the forward propagation (FP) and backward propagation (BP) processes between the client server and the main server are systematically divided into five distinct steps to ensure efficient communication and computation, optimizing the distributed learning process.

\section{Evaluation and Results Analysis}
\label{evaluation}
\subsection{Prototype Implementation}
We implement our \textit{SplitFT} atop of \textit{PyTorch} and \textit{Flower}. We design the fine-tuning framework as Fig. \ref{fig1}. \textit{Local FedAvg Server} is utilized to aggregate the model for the layers that are trained on clients. \textit{Main Server} is designed to train the other layers of the model besides the layer on the clients. 

\begin{figure*}[!ht]
    \centering
    \subfloat[Different LoRA ranks on Cutlayers.]{\includegraphics[width=0.33\linewidth, height=0.21\linewidth]{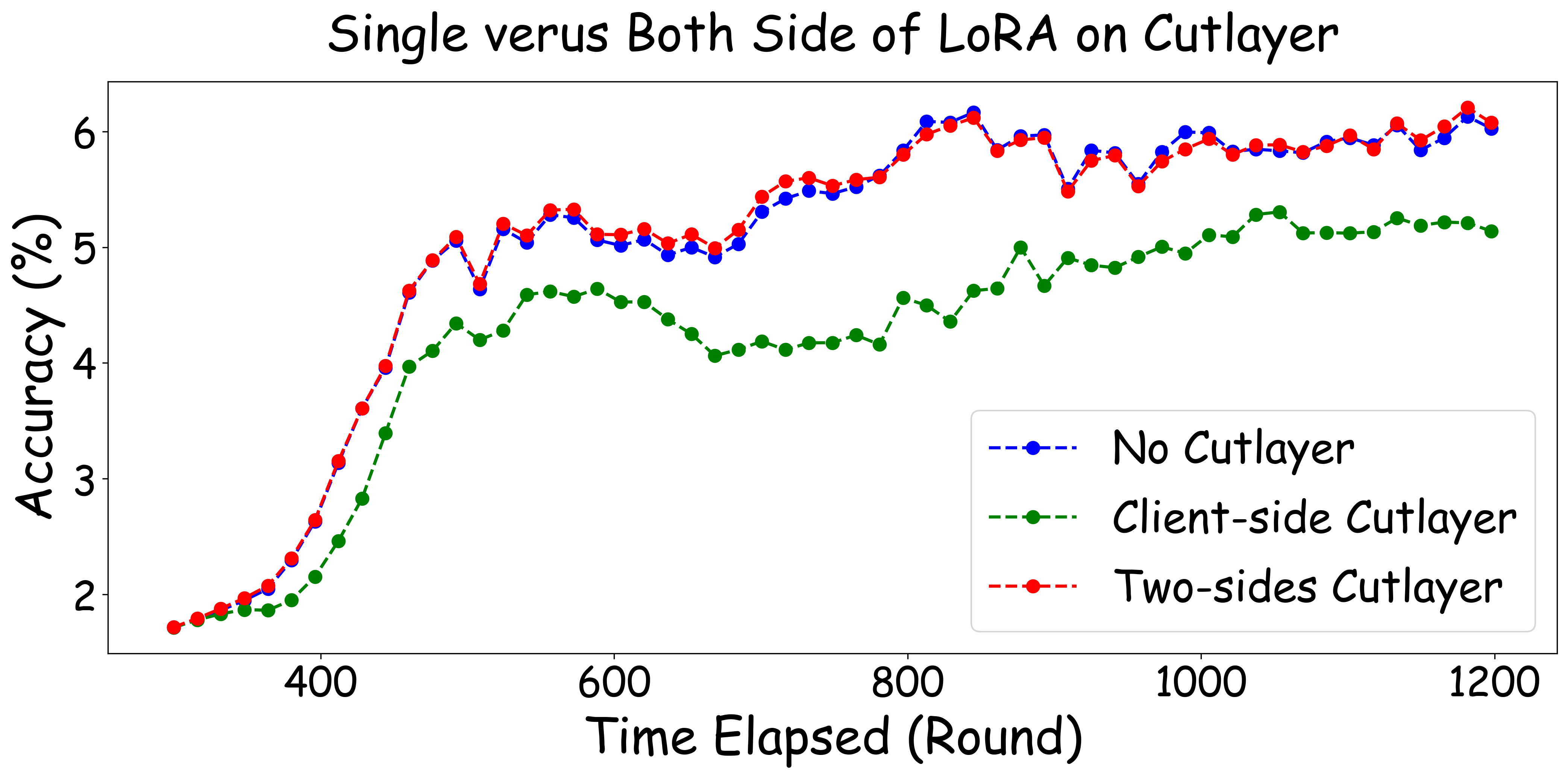}} 
    \subfloat[Different Cutlayer Settings on Baseline.]{\includegraphics[width=0.33\linewidth, height=0.21\linewidth]{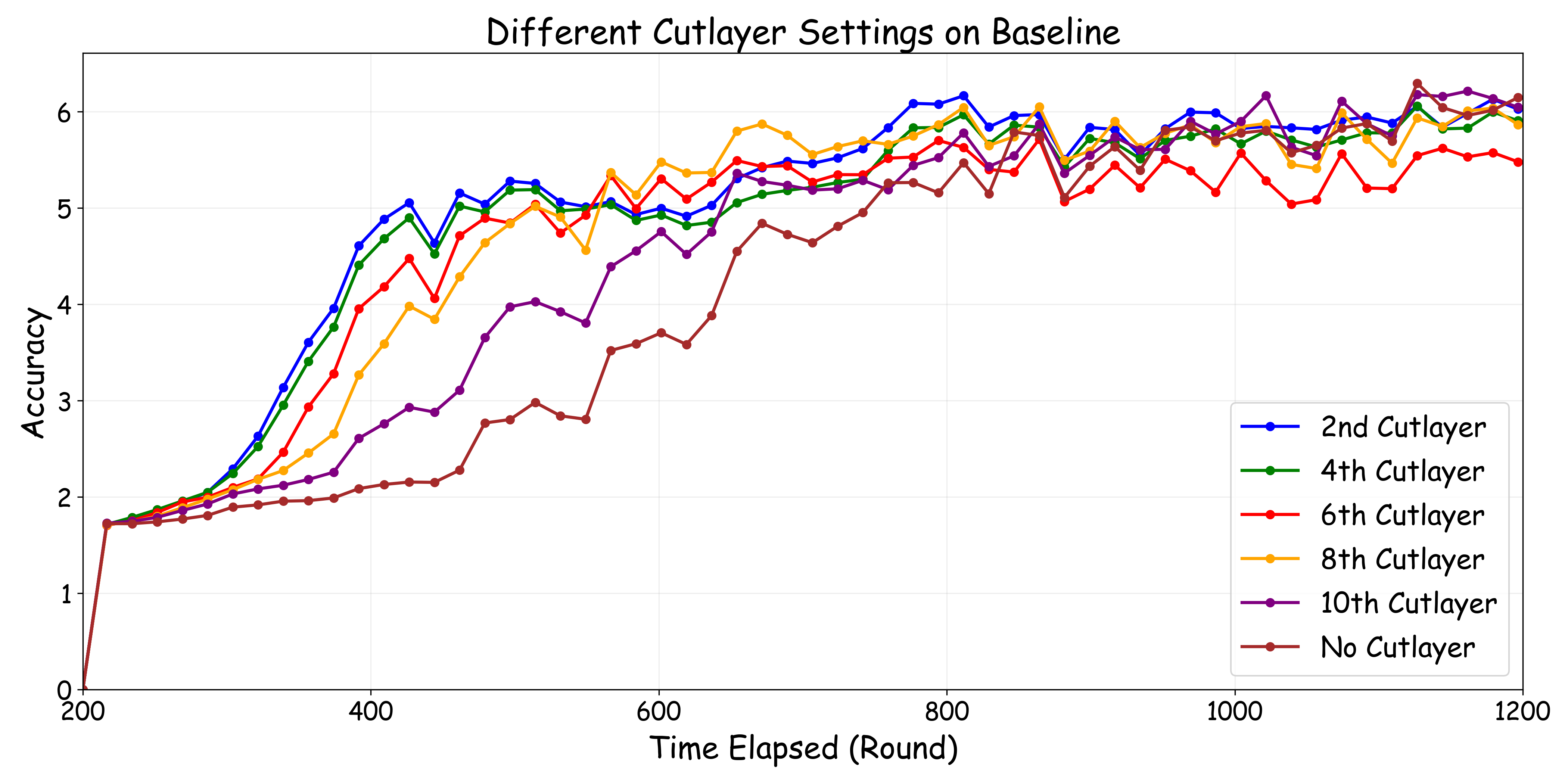}} 
    \subfloat[Same Cutlayer \& Different LoRA Ranks.]{\includegraphics[width=0.33\linewidth, height=0.21\linewidth]{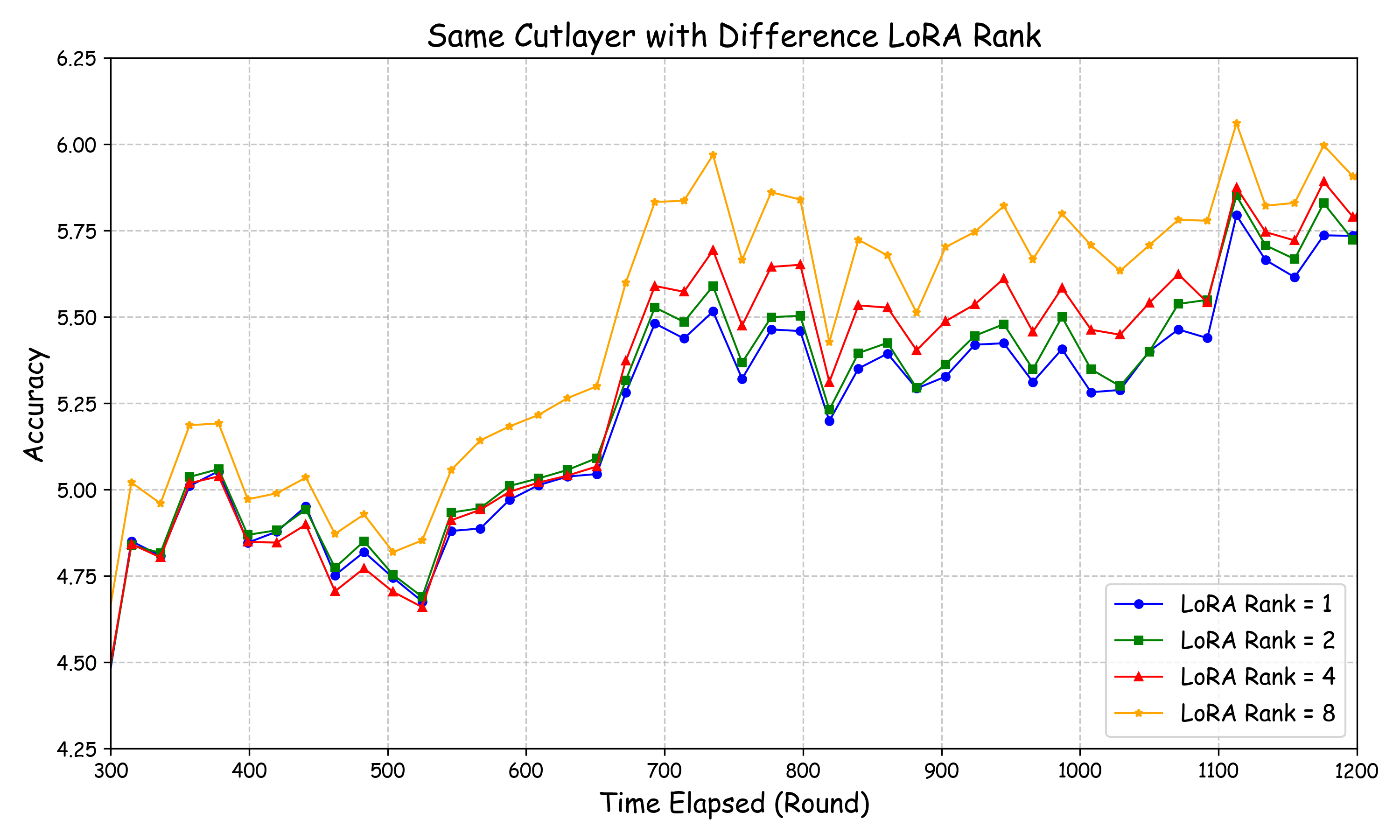}} 
    \caption{The Impact of LoRA Rank and Cutlayer on Model Performance and Quality.}
    \label{fig2}
\end{figure*}

\subsection{Evaluation Methodology}

\noindent \textbf{\textit{Testbed.}} The experiments were conducted on NVIDIA GeForce RTX 3090 GPUs using PyTorch 2.3. 

\noindent \textbf{\textit{Benchmarks.}} We choose 3 most popular LLMs to evaluate the system performance of \textit{SplitFT}, including \ding{172} GPT2-small, \ding{173} OPT-125M, and \ding{174} GPT-Neo 125M. These models are evaluated on the Wikitext2-v1 dataset, which contains 36,700 training samples, 3,760 validation samples, and 4,360 test samples extracted from Wikipedia articles. 

\noindent \textbf{\textit{Experimental Hyper-Parameter Setup.}} Key fine-tuning hyper-parameters included a batch size of 4, learning rate of $5 \times 10^{-5}$, maximum sequence length of 512 tokens, and local data partitions of 12,000 samples per client. LoRA configuration used rank $r_{\text{cut}}=8$ for cutlayer attention modules and $r_{\text{others}}=16$ for other layers. The baseline configuration maintained identical client-server layer splits (first 2 GPT2Blocks on clients, remaining 10 on server) with matching LoRA settings for comparative analysis.

\noindent \textbf{\textit{Baselines.}} We evaluate SplitFT on the Wikitext2-v1 dataset for next-sentence generation, comparing against a baseline with fixed split layers. Our experiments use GPT2-small as the base model, with validation on OPT-125M and GPT-Neo 125M to demonstrate generalizability. The framework is implemented with 5 clients, where each client processes the first 2 GPT2Blocks and the server handles the remaining 10 layers. Key configurations include LoRA adaptation on attention modules with rank reduction at cutlayers.

\subsection{Summary of Experimental Results}

The experimental results validate the effectiveness of the proposed SplitFT framework across multiple dimensions, including data heterogeneity, LoRA rank configurations, splitting strategies, and model architectures. By leveraging adaptive layer adjustment and parameter-efficient fine-tuning, SplitFT achieves state-of-the-art performance while maintaining computational efficiency, particularly in federated and distributed learning scenarios. Key findings include:

$\bullet$ \textbf{LoRA Rank and Cutlayer Configuration}: Properly configuring the LoRA ranks and cutlayer significantly impacts model performance and training efficiency. Symmetric adjustments to the LoRA ranks on both client and server sides yield better performance compared to asymmetric configurations.

$\bullet$ \textbf{Adaptive Layer Splitting}: The \textit{SplitFT} method enhances model accuracy and convergence speed while reducing computational overhead for clients with lower-quality data, outperforming the \textit{Same Split} baseline.

$\bullet$ \textbf{Data Distribution Robustness}: SplitFT maintains superior performance in Non-IID data distributions, effectively adapting to heterogeneous client environments.

$\bullet$ \textbf{Model Generalizability}: The framework demonstrates consistent performance improvements across different language models, highlighting its adaptability and robustness.

\subsection{Impact of LoRA Ranks and Cutlayer}

\subsubsection{The Effect of LoRA Rank Reduction of Cutlayer on Model Performance}
To evaluate the effectiveness of our proposed approach to reduce the LoRA rank for cutlayer to reduce the communication overhead during the fine-tuning procedure, we compare three different kinds of cases based on the GPT2-small model (12 blocks) and show the results in Fig. \ref{fig2}(a). In Fig. \ref{fig2}(a), \ding{182} \textit{No Cutlayer} refers to fine-tuning the model (GPT2-small) and setting the LoRA rank for each block for all layers as 16; \ding{183} \textit{Client-side Cutlayer} refers to fine-tuning the model (GPT2-small) with the setting that reduces the LoRA rank for the cutlayer of client-side as 8, while the LoRA rank for all other layers is 16; \ding{184} \textit{Two-side Cutlayer} refers to fine-tuning the model (GPT2-small) with the setting that reduces the LoRA rank for culayer of client-side and server-side as 8, while the LoRA rank for all other layers is 16. We set the cutlayer as the second layer of the model (GPT2-small) for all three cases above.

From Fig. \ref{fig2}(a), we observe that adjusting the LoRA rank on both sides of the cutlayer is critical for balancing feature extraction and gradient flow. A mismatched rank can adversely impact performance, whereas symmetric or collaboratively tuned rank adjustments can achieve optimal results. As illustrated in Fig.~\ref{fig2}(a), when all layers' LoRA ranks are set to 16, reducing the LoRA rank on only one side of the cutlayer does not yield better performance compared to jointly decreasing the LoRA rank on both sides. We systematically analyzed the impact of different LoRA rank configurations by varying \( r_{\text{cut}} \) and \( r_{\text{others}} \). The results demonstrate that setting \( r_{\text{cut}} \ll r_{\text{others}} \) optimizes both computational efficiency and model performance. Additionally, the adaptive layer adjustment mechanism in \textit{SplitFT} effectively balances workload distribution across clients while maintaining competitive accuracy, regardless of differences in LoRA ranks.

\begin{figure*}[!ht]
    \centering
    \subfloat[Adaptive SplitFT v.s. Same Split.]{\includegraphics[width=0.33\linewidth, height=0.2\linewidth]{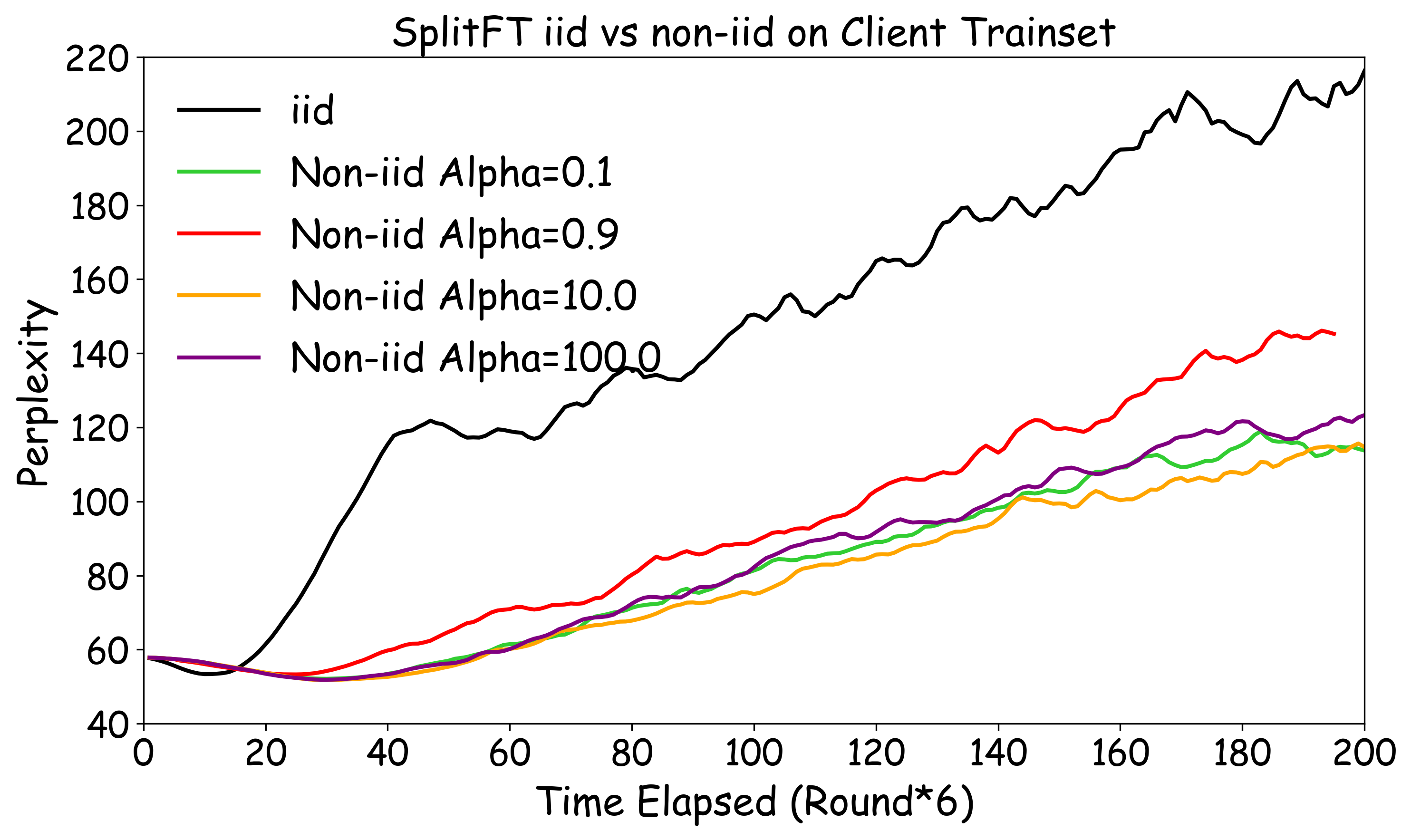}} 
    \subfloat[Baseline versus SplitFT on Client Trainset.]{\includegraphics[width=0.33\linewidth, height=0.2\linewidth]{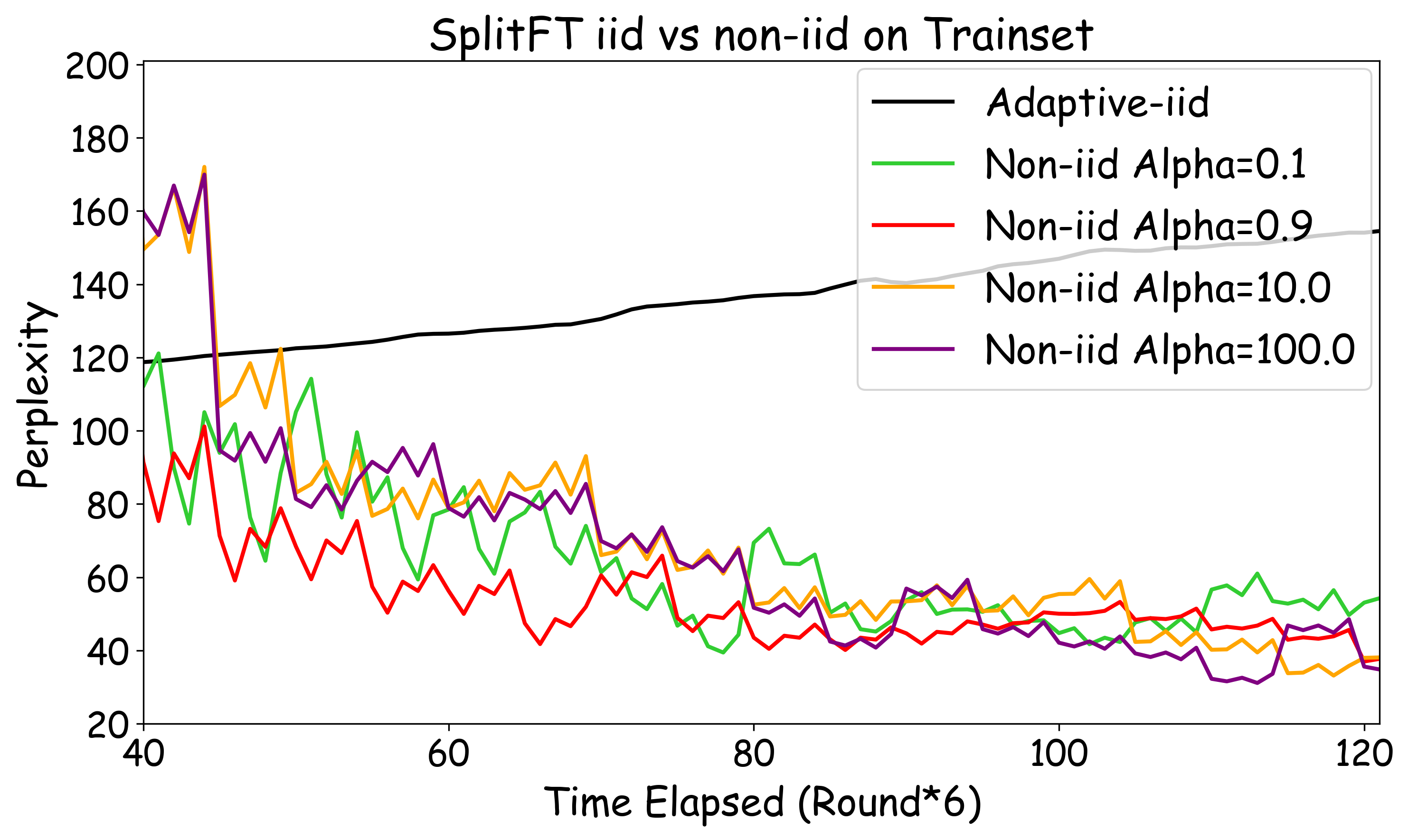}} 
    \subfloat[Baseline versus SplitFT on Client Trainset.]{\includegraphics[width=0.33\linewidth, height=0.2\linewidth]{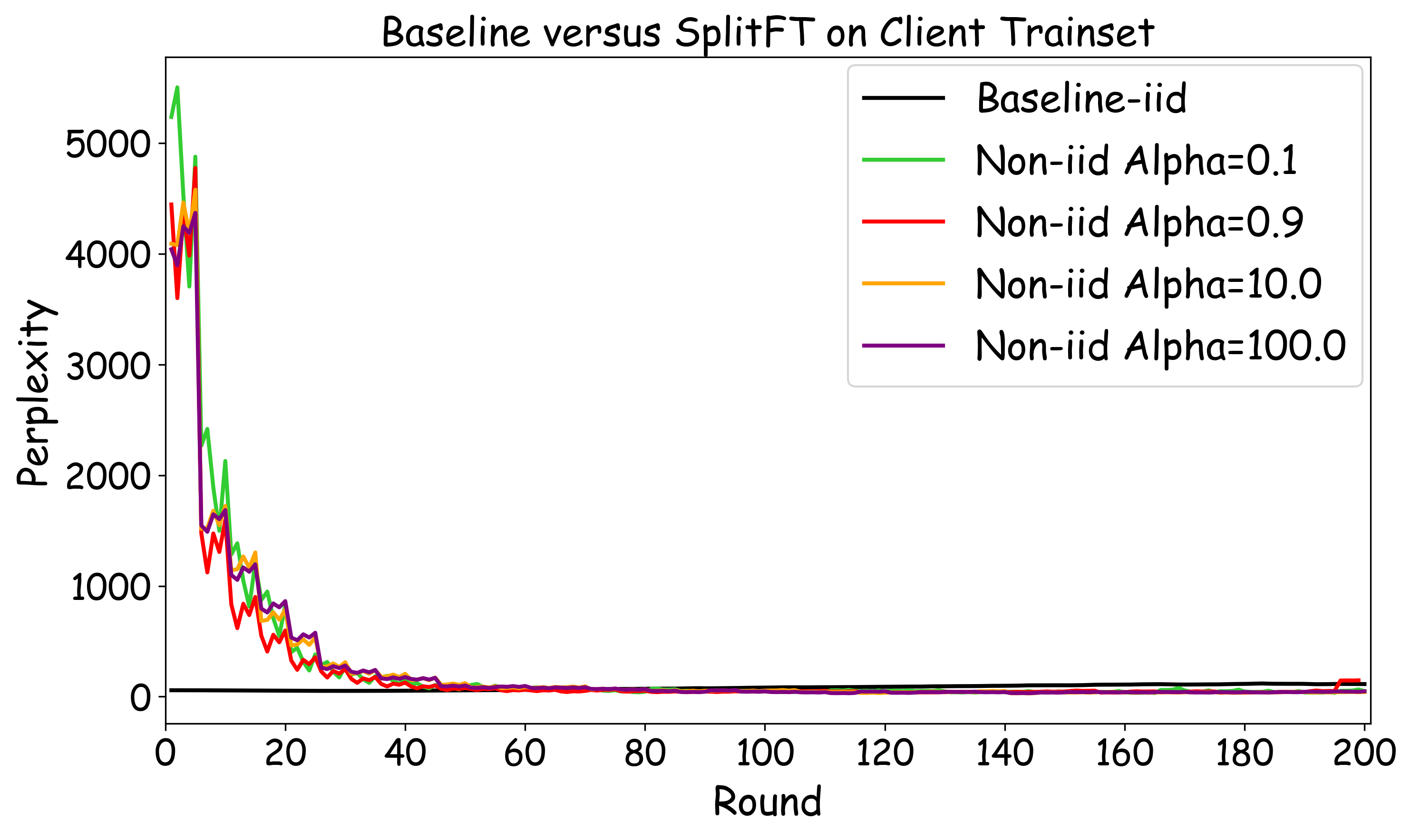}}
    \caption{The Performance Comparison for \textit{SplitFT} and Baselines.}
    \label{fig5}
\end{figure*}

\begin{table*}[!ht]
\centering
\caption{Comparison for Accuracy, Elapsed Time, Round Time, and Communication Overhead for Different Cutlayers.}
    \centering
    \begin{tabular}{c|cccc}
    \toprule[1pt]
    \textbf{Cutlayer} & \textbf{Max Accuracy} & \textbf{Mean Elapsed Time (s)} & \textbf{Mean Round Time (s)} & \textbf{Max Comm Overhead (MB)} \\
    \midrule[0.8pt]
    2 & 0.0606 & 810.4379 & 0.0347 & 3475.3674 \\
    4 & 0.0571 & 863.2450 & 0.0424 & 3534.3875 \\
    6 & 0.0605 & 934.1831 & 0.0547 & 3593.4076 \\
    8 & 0.0621 & 1113.3617 & 0.0635 & 3652.4278 \\
    10 & 0.0629 & 1104.7156 & 0.0759 & 3711.4478 \\
    No Cut & 0.0617 & 6897.3059 & 0.0404 & 3490.0521 \\
    \bottomrule[1pt]
    \end{tabular}

    \label{tab1}
\end{table*}

\begin{table*}[!t]
    \centering
    \caption{Comparison for Accuracy, Elapsed Time, Round Time, and Communication Overhead for Different LoRA Ranks.}
    \begin{tabular}{c|ccccc}
    \toprule[1pt]
    \textbf{Rank} & \textbf{Max Acc.} & \textbf{Elapsed Time (s)} & \textbf{Round Time (s)} & \textbf{Comm Overhead (MB)} & \textbf{Trainable Param.} \\
    \midrule[0.8pt]
    1 & 0.0579 & 2277.5402 & 0.0558 & 3462.5080 & 0.08M \\
    2 & 0.0585 & 2138.1923 & 0.0389 & 3464.3450 & 0.015M \\
    4 & 0.0589 & 2293.3771 & 0.0393 & 3468.0193 & 0.031M \\
    8 & 0.0606 & 1597.7380 & 0.0347 & 3475.3674 & 0.062M \\
    \bottomrule[1pt]
    \end{tabular}
    \label{tab2}
\end{table*}

\subsubsection{The Effect of Cutlayer on Model Performance}

To further evaluate the effectiveness of the cutlayer for the model performance, we conduct the experiment based on the GPT2-small model and set different cutlayers, and show results in Fig. \ref{fig2}(b), where we set the LoRA rank as 8 and set the cutlayer as $2,4,6,8,10$. We also list the best accuracy for each case in Tab. \ref{tab1}. From Fig. \ref{fig2}(b) and Tab. \ref{tab1}, we observe that positioning the cutlayer earlier in the model reduces the computational load on the client, thereby facilitating the participation of lightweight devices. Transferring more workload to the server enhances gradient flow and enables the optimization process to converge more rapidly, thereby improving training performance. However, this approach also increases communication overhead and imposes greater demands on the server.

\subsubsection{The Effect of LoRA Rank on Model Performance}
To assess the impact of LoRA rank on model performance, we conducted experiments using the GPT2-small model, setting the cutlayer at the second layer with varying LoRA ranks of $1,2,4,8$ for the cutlayer. The LoRA rank for all other layers, excluding the cutlayer, was set to 16. The results are presented in Fig. \ref{fig2}(c), and the best accuracy for these scenarios is listed in Tab. \ref{tab2}. From Fig. \ref{fig2}(c) and Tab. \ref{tab2}, we observe that the convergence rate and final accuracy remain nearly consistent across different LoRA ranks, while higher ranks can expedite the fine-tuning process. Smaller ranks effectively reduce communication overhead and alleviate the demands on computational resources without compromising model performance.

\begin{figure*}[!h]
    \centering
    \subfloat[SplitFT on OPT-125M 0-1200 Rounds]{\includegraphics[width=0.32\linewidth, height=0.2\linewidth]{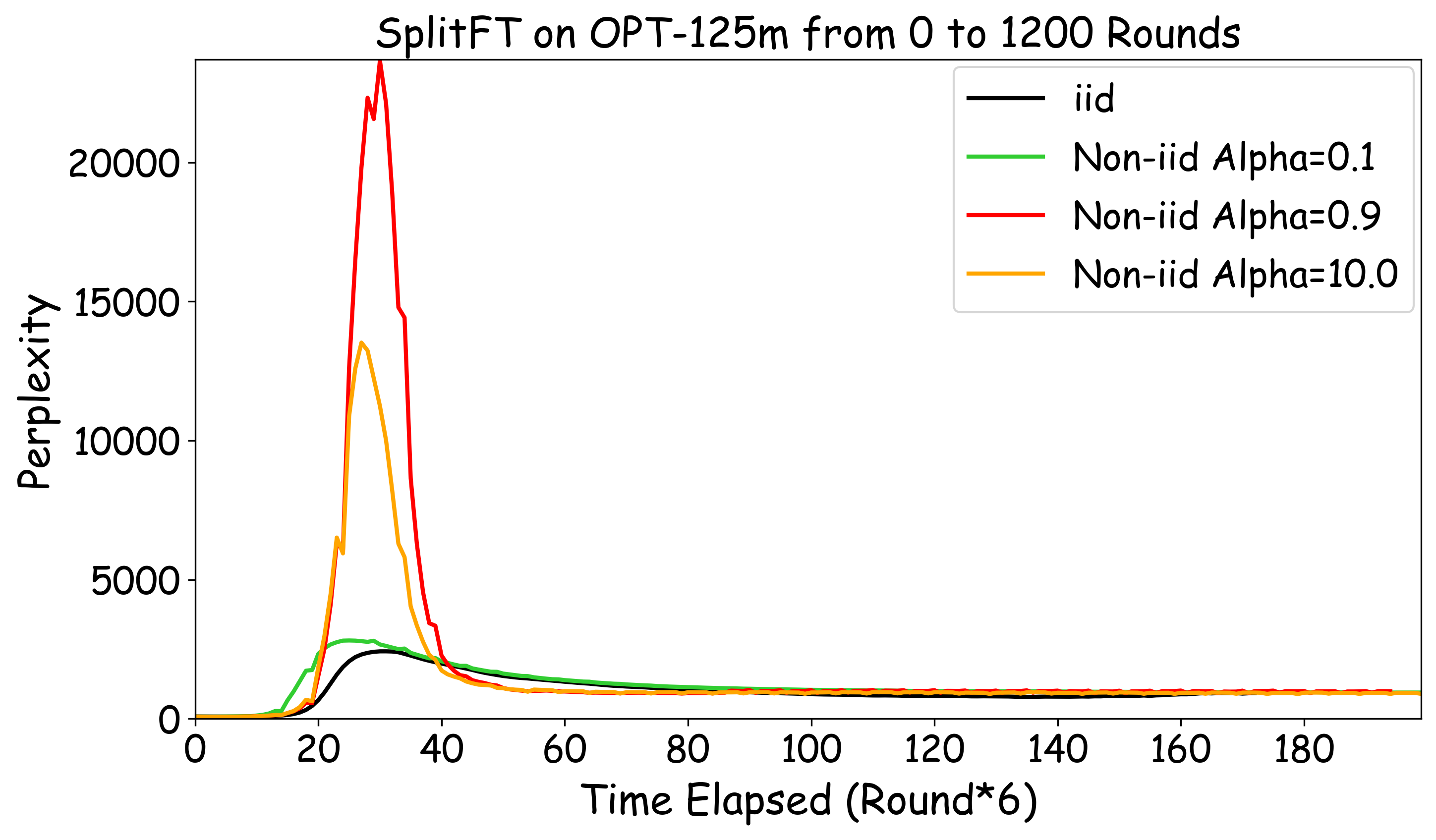}} 
    \subfloat[SplitFT on OPT-125M 300-1200 Rounds]{\includegraphics[width=0.32\linewidth, height=0.2\linewidth]{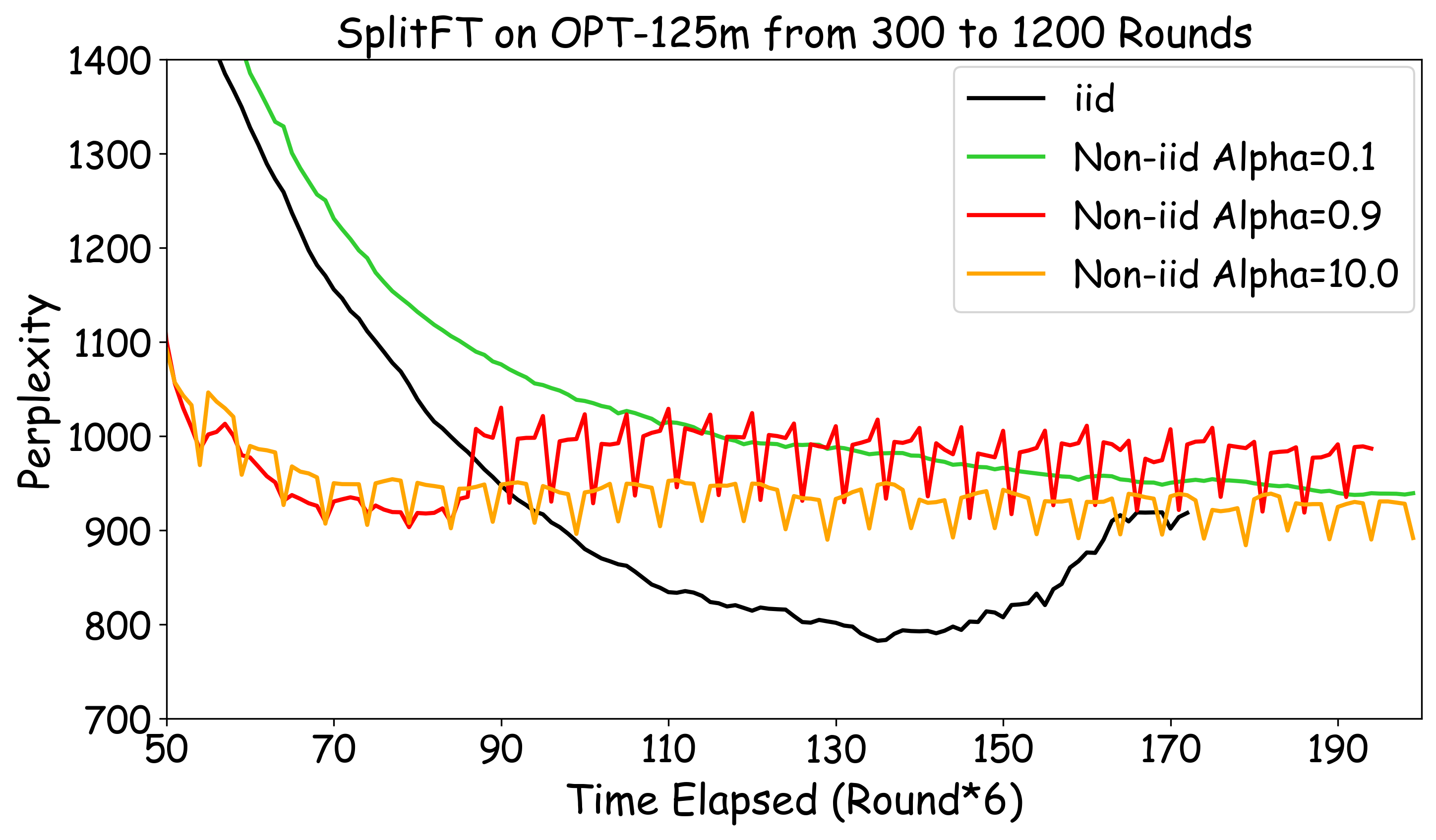}} 
    \subfloat[SplitFT on GPT-Neo]{\includegraphics[width=0.32\linewidth, height=0.2\linewidth]{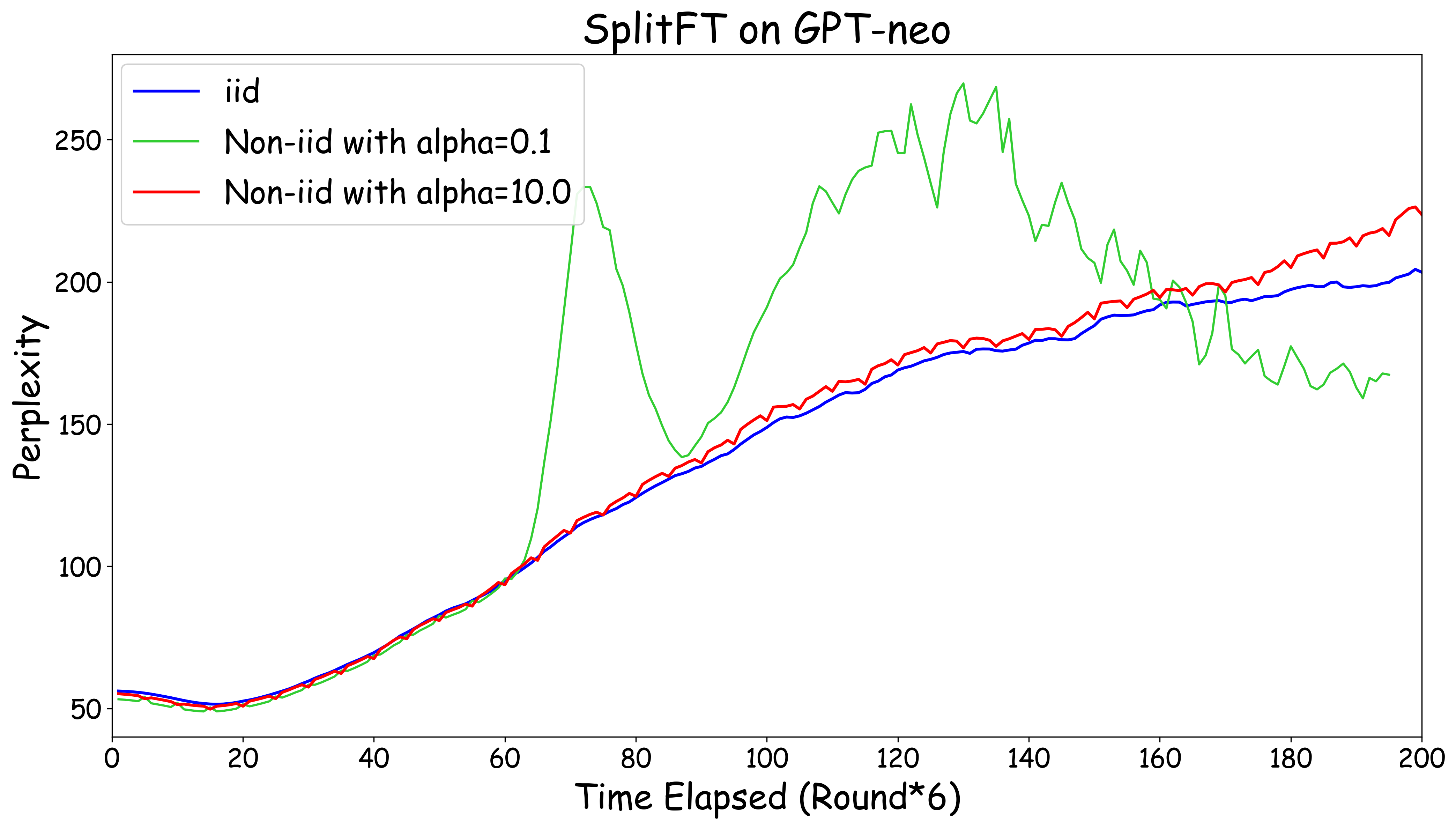}} 
    \caption{Generalizability of SplitFT Across Different Models}
    \label{fig8}
    \vspace{-6pt}
\end{figure*}

\subsection{Comparison of \textit{SplitFT} and  Baselines}
\subsubsection{The comparison for adaptive \textit{SplitFT} and same \textit{SplitFT}}
To further evaluate the effectiveness of our adaptive cutlayer strategy, we compare our adaptive \textit{SplitFT} to the baseline with the same cutlayer across all clients in Fig. \ref{fig5}(a). In Fig. \ref{fig5}(a), the label \textit{iid} refers to the setting that the $cutlyaer=2$ for all clients and the LoRA rank for cutlyaer is 8, and the LoRA rank for all other layers is 16. The label \textit{non-iid Alpha$=x$} refers to the setting that we adjust the cutlayer adaptively during the fine-tuning procedure, and set the LoRA rank for cutlayer as 8 while setting the LoRA rank for all other layers as 16, where $Alpha = 0.1, 0.9, 10.0, 100.0$ refers to the hyperparameters $\alpha$ for the Dirichlet approach.

From Fig. \ref{fig5}(a), we observe that although the baseline initially fits more rapidly, its swift pace complicates the identification of whether a point represents the minimum fit, leading to overfitting issues. In contrast, our method, despite fitting slightly slower, yields a model with lower perplexity than the same split method and subsequently rises gradually, facilitating the quick determination of the fitting point. The comparative analysis further demonstrates that \textit{SplitFT} not only improves perplexity and convergence speed but also reduces computational overhead for clients with lower-quality data. The results indicate that the baseline's final perplexity on IID datasets is significantly higher compared to other \textit{SplitFT} configurations. Conversely, when employing \textit{SplitFT}, the framework swiftly adapts to an optimal cutlayer location based on the dataset's characteristics, converging at a lower perplexity. This suggests that the auto-adaptation mechanism for adjusting the number of layers on the client effectively enhances model performance. Moreover, \textit{SplitFT} consistently achieves superior performance in Non-IID settings compared to the Same Split method, underscoring its ability to adapt to heterogeneous data distributions.

\subsubsection{The Effect of Distribution of Dataset on Model Performance}

To assess the impact of training dataset distribution on model performance, we conducted experiments using the GPT2-small model across five different settings, as illustrated in Fig. \ref{fig5}(b) and Fig. \ref{fig5}(c). In these figures, we adaptively adjusted the cutlayer during the fine-tuning process, setting the LoRA rank for the cutlayer at 8, while all other layers were set at 16. We compared model performance, measured by perplexity, across various training dataset distributions, including IID and Non-IID settings with different values of $\alpha = 0.1, 0.9, 10.0, 100.0$. From Fig. \ref{fig5}(b) and Fig. \ref{fig5}(c), it is evident that in IID scenarios, test perplexity increases with training rounds due to overfitting to the uniform data distribution and over-smoothing during global aggregation, which diminishes the model's generalization ability to unseen test data. Conversely, in Non-IID scenarios, the training perplexity of clients decreases with training rounds, attributed to personalized optimization for heterogeneous data distributions, \textit{SplitFT}'s dynamic layer and LoRA rank adjustments that enhance adaptation to client-specific data, and reduced gradient conflicts during global aggregation, facilitating more efficient local optimization. These factors elucidate why the baseline model and \textit{SplitFT} exhibit opposite trends in their behavior.

To evaluate the robustness of \textit{SplitFT} in handling Non-IID scenarios, we conducted experiments using both IID and Non-IID data distributions. In the Non-IID settings, the degree of heterogeneity was controlled using the Dirichlet parameter \( \alpha \), where smaller \( \alpha \) values indicate highly skewed distributions, and larger \( \alpha \) values approximate IID-like scenarios. Our findings revealed that with the lowest \( \alpha = 0.1 \), representing highly heterogeneous data, our framework maintained similar or slightly lower perplexity after 1200 rounds compared to larger \( \alpha \) values (\( \alpha = \{10, 100\} \)), which correspond to more balanced data distributions. These results demonstrate that \textit{SplitFT} consistently achieves superior performance in Non-IID settings compared to the Same Split method, underscoring its ability to adapt effectively to heterogeneous data distributions.

\subsection{Generalizability Across Models}
To evaluate the generalizability of our proposed \textit{SplitFT}, we conduct a comprehensive experiment based on three of the most popular language models with varying architectures and scales, including GPT2-small, OPT-125M, and GPT-Neo 125 M. Across all models, \textit{SplitFT} consistently outperformed the baseline illustrated in Fig. \ref{fig8} in both perplexity and efficiency, demonstrating its adaptability to different model architectures. In Fig.~\ref{fig8} (a) (b) (c), \textit{IID} label refers to that we fine-tune the model based on the IID dataset and adjust the cutlayer adaptively during the fine-tuning procedure, and set the LoRA rank for the cutlayer as 8 while all other layers as 16. \textit{Non-IID} refers to that we compare the model performance (perplexity) for different distributions of the training dataset, including the Non-IID settings with different $\alpha = 0.1, 0.9, 10.0, 100.0$. From Fig. \ref{fig8}, we observe that \textit{SplitFT} can achieve the same model performance and quality on both IID and Non-IID environments across various popular models, which further highlights the robustness and generalization ability of our proposed \textit{SplitFT}.

\section{Related Works}
\label{related_works}

\textit{Federated Split Learning.}
Federated Split Learning is a paradigm combining the advantages of Federated Learning\cite{mcmahan2023communicationefficientlearningdeepnetworks} and Split Learning\cite{vepakomma2018splitlearninghealthdistributed}, which aims at reducing the computational burden on the clients while protecting data privacy \cite{zhang2024theoreticalanalysisprivacyleakage, 10.1007/978-981-99-9785-5_18, rafi2023fairnessprivacypreservingfederatedlearning}. Recently, various Federated Split Learning models and architectures have been proposed, highlighting basic structures, communication methods, and adaptability to data distribution \cite{R1}. To further address the problems of high communication overhead and server storage requirements incurred by the linearly growing number of clients in traditional FSL, CSE-FSL\cite{R2} is proposed by introducing an auxiliary network to reduce overhead by updating submodels locally at the client using auxiliary network and keeping only a single global model at the server. CSE-FSL also deducts the frequency of sending smashed data to save communication costs. Luo et al. proposed a Federated Split learning framework via Mutual Knowledge Distillation (FSMKD)\cite{R3}, which integrated FL with SL in a two-way manner. FSMKD can therefore support personalized local models while leveraging information from diverse heterogeneous learning tasks to enhance global model training via deep mutual learning.

Federated Split Learning combines federated learning's parallel training \cite{di2025survey, chenfeddes, liu2025ocelot, huang2024optimized, huang2023c, huang2025zccl,ma2025compression,ye2025efficient,ngcan,wei2025dynamic,hu2025fastrei,xu2026deepebc} with split learning's model splitting strategy. They design a two-body structure, including a head:personalized-local-body: tail network as a local model and a head:shared-server-body: tail network as a global model. Various Federated Split Learning models and architectures have been proposed, highlighting basic structures, communication methods, and adaptability to data distribution \cite{R1}.

Compared with FSL, FL trains full foundation models on each client based on local data accordingly and uploads model parameters or gradients to the server for aggregation to form a global model. Although FL is significant in that it utilizes clients' computational resources efficiently with no need to transmit raw data centrally, training full models on clients is a heavy burden on resource-constrained devices, with potential exposures to data privacy leakage due to access to full models by both clients and the server. On the other hand, the SL user serial training method is that after one client completes forward propagation, the server performs backward propagation and returns gradients, which reduces training efficiency when there are many clients.

SplitFed\cite{R1} is a structure where each client and server retain part of the model. Clients send intermediate activations to the server after forward propagation, the server completes backward propagation and returns gradients to clients; all clients execute this process in parallel, then the server aggregates client sub-model parameters from all clients. It provides better model privacy than FL, faster training speed than SL, and similar accuracy and communication efficiency. 

To further address the problems of high communication overhead and server storage requirements incurred by the linearly growing number of clients in traditional FSL, CSE-FSL\cite{R2} is proposed by introducing an auxiliary network to reduce overhead by updating submodels locally at the client using the auxiliary network and keeping only a single global model at the server. CSE-FSL also deducts the frequency of sending smashed data to save communication costs.

Luo et al. proposed a Federated Split learning framework via Mutual Knowledge Distillation (FSMKD)\cite{R3}, which integrated FL with SL in a two-way manner. They design a two-body structure, including the head:personalized-local-body: tail network as the local model and the head:shared-server-body: tail network as the global model. FSMKD can therefore support personalized local models while leveraging information from diverse heterogeneous learning tasks to enhance global model training via deep mutual learning.

\noindent \textit{Parameter-Efficient Fine-Tuning (PEFT).} Low Rank Adaptation\cite{hu2021loralowrankadaptationlarge} is proposed to reparameterize the weight matrices in the transformer series models in low rank and train two small matrices while freezing the whole original model weights. Experiments show that LoRA can reduce the trainable parameters by a factor of approximately 10,000 and a reduction in memory by a factor of 3 in models such as DeBERTa, GPT-2/3, etc., with comparable model performance to full fine-tuning. Later improvement in LoRA, like LoRA-Pro\cite{R4}, introduces equivalent gradient constraints to narrow the gap with full fine-tuning.  Others, such as Liu et al., proposed DoRA\cite{R5} to incorporate learnable scaling vectors, Kalajdzievski et al. proposed rsLoRA\cite{R6} to improve gradient stabilization, and Hayou et al. proposed various variants such as LoRA+\cite{R7} to explore a more flexible low-rank updating. Li et.al proposed Prefix Tuning\cite{R8}. It inserts a sequence of learnable consecutive prefix vectors into each layer of the transformer.  Houlsby et al.\cite{R9} first proposed adapter modules that insert bottleneck structures after each converter sublayer. These modules contain downsampling-nonlinear-upsampling layers that effectively add a small number of parameters to each network layer. Experiments have shown that on BERT-Large, the performance of Adapter tuning is only ~0.4\% lower than full fine-tuning on benchmarks such as GLUE, while adding only about 3\% more trainable parameters. Various parameter-efficient adapter variants, such as Hyperformer and Compacter, have been proposed to reduce the number of parameters or improve performance.

\vspace{6pt}
\section{Conclusion}
\label{conclusion}
We propose \textit{SplitFT}, a robustness system designed for LLMs fine-tuning in federated learning environments, enabling different clients to set different cut-off criteria according to their computation resources and trained model performance, thus improving the overall system performance for LLMs fine-tuning. \textit{SplitFT} also proposes to reduce the LoRA rank in cutlayer to reduce the communication overhead. In addition to simulating the heterogeneous data in real-world applications for our proposed split federated learning system, \textit{SplitFT} proposes a length-based Dirichlet approach to divide the training data into different clients. Extensive experimental results show that our proposed approach outperforms the state-of-the-art approach for fine-tuning time efficiency and model performance based on various popular benchmarks. The \textit{SplitFT} demonstrated adaptability across diverse data distributions (IID and Non-IID) and scalability with different LoRA ranks and cutlayer strategies.

\bibliographystyle{IEEEtran}
\bibliography{CCGrid_2026}

\begin{thebibliography}{10}
\providecommand{\url}[1]{#1}
\csname url@samestyle\endcsname
\providecommand{\newblock}{\relax}
\providecommand{\bibinfo}[2]{#2}
\providecommand{\BIBentrySTDinterwordspacing}{\spaceskip=0pt\relax}
\providecommand{\BIBentryALTinterwordstretchfactor}{4}
\providecommand{\BIBentryALTinterwordspacing}{\spaceskip=\fontdimen2\font plus
\BIBentryALTinterwordstretchfactor\fontdimen3\font minus \fontdimen4\font\relax}
\providecommand{\BIBforeignlanguage}[2]{{%
\expandafter\ifx\csname l@#1\endcsname\relax
\typeout{** WARNING: IEEEtran.bst: No hyphenation pattern has been}%
\typeout{** loaded for the language `#1'. Using the pattern for}%
\typeout{** the default language instead.}%
\else
\language=\csname l@#1\endcsname
\fi
#2}}
\providecommand{\BIBdecl}{\relax}
\BIBdecl

\bibitem{li2024scopingreviewusinglarge}
\BIBentryALTinterwordspacing
L.~Li, J.~Zhou, Z.~Gao, W.~Hua, L.~Fan, H.~Yu, L.~Hagen, Y.~Zhang, T.~L. Assimes, L.~Hemphill, and S.~Ma, ``A scoping review of using large language models (llms) to investigate electronic health records (ehrs),'' 2024. [Online]. Available: \url{https://arxiv.org/abs/2405.03066}
\BIBentrySTDinterwordspacing

\bibitem{wangdan}
\BIBentryALTinterwordspacing
D.~Wang, B.~Liu, R.~Lu, Z.~Zhang, and S.~Zhu, ``Storellm: Energy efficient large language model inference with permanently pre-stored attention matrices,'' in \emph{Proceedings of the 16th ACM International Conference on Future and Sustainable Energy Systems}, ser. E-Energy '25.\hskip 1em plus 0.5em minus 0.4em\relax New York, NY, USA: Association for Computing Machinery, 2025, p. 398–406. [Online]. Available: \url{https://doi.org/10.1145/3679240.3734604}
\BIBentrySTDinterwordspacing

\bibitem{ma2025moe}
S.~Ma, Z.~Zhang, S.~Di, B.~Liu, X.~Yu, X.~Lu, and D.~Wang, ``Moe-compression: How the compression error of experts affects the inference accuracy of moe model?'' \emph{arXiv preprint arXiv:2509.07727}, 2025.

\bibitem{ma2025compression}
------, ``Compression error sensitivity analysis for different experts in moe model inference,'' in \emph{Proceedings of the SC'25 Workshops of the International Conference for High Performance Computing, Networking, Storage and Analysis}, 2025, pp. 339--348.

\bibitem{openai2024gpt4technicalreport}
\BIBentryALTinterwordspacing
O.~Group, ``Gpt-4 technical report,'' 2024. [Online]. Available: \url{https://arxiv.org/abs/2303.08774}
\BIBentrySTDinterwordspacing

\bibitem{c:1}
\BIBentryALTinterwordspacing
J.~Kaddour, J.~Harris, M.~Mozes, H.~Bradley, R.~Raileanu, and R.~McHardy, ``Challenges and applications of large language models,'' 2023. [Online]. Available: \url{https://arxiv.org/abs/2307.10169}
\BIBentrySTDinterwordspacing

\bibitem{vaswani2023attentionneed}
\BIBentryALTinterwordspacing
A.~Vaswani, N.~Shazeer, N.~Parmar, J.~Uszkoreit, L.~Jones, A.~N. Gomez, L.~Kaiser, and I.~Polosukhin, ``Attention is all you need,'' 2023. [Online]. Available: \url{https://arxiv.org/abs/1706.03762}
\BIBentrySTDinterwordspacing

\bibitem{c:2}
\BIBentryALTinterwordspacing
S.~Garg, T.~Vu, and A.~Moschitti, ``Tanda: Transfer and adapt pre-trained transformer models for answer sentence selection,'' 2019. [Online]. Available: \url{https://arxiv.org/abs/1911.04118}
\BIBentrySTDinterwordspacing

\bibitem{informatics11030057}
\BIBentryALTinterwordspacing
Z.~A. Nazi and W.~Peng, ``Large language models in healthcare and medical domain: A review,'' \emph{Informatics}, vol.~11, no.~3, 2024. [Online]. Available: \url{https://www.mdpi.com/2227-9709/11/3/57}
\BIBentrySTDinterwordspacing

\bibitem{peft}
S.~Mangrulkar, S.~Gugger, L.~Debut, Y.~Belkada, S.~Paul, and B.~Bossan, ``Peft: State-of-the-art parameter-efficient fine-tuning methods,'' in \emph{Peft: State-of-the-art parameter-efficient fine-tuning methods}, 2022.

\bibitem{liu2025hlora}
Q.~Liu, Z.~Zhang, X.~Yao, and B.~Liu, ``Hlora: Efficient federated learning system for llm heterogeneous fine-tuning,'' \emph{arXiv preprint arXiv:2503.00813}, 2025.

\bibitem{zhang2025cllora}
P.~Zhang, Z.~Zhang, S.~Di, Y.~Xin, and B.~Liu, ``Cllora: An approach to measure the effects of the context length for llm fine-tuning,'' \emph{arXiv preprint arXiv:2502.18910}, 2025.

\bibitem{Lester2021}
B.~Lester, R.~Al-Rfou, and N.~Constant, ``The power of scale for parameter-efficient prompt tuning,'' \emph{arXiv preprint arXiv:2104.08691}, 2021.

\bibitem{Houlsby2019}
N.~Houlsby, A.~Giurgiu, S.~Jastrzebski, B.~Morrone, Q.~De~Laroussilhe, A.~Gesmundo, M.~Attariyan, and S.~Gelly, ``Parameter-efficient transfer learning for nlp,'' in \emph{International conference on machine learning}.\hskip 1em plus 0.5em minus 0.4em\relax PMLR, 2019, pp. 2790--2799.

\bibitem{hu2022lora}
E.~J. Hu, Y.~Shen, P.~Wallis, Z.~Allen-Zhu, Y.~Li, S.~Wang, L.~Wang, W.~Chen \emph{et~al.}, ``Lora: Low-rank adaptation of large language models.'' \emph{ICLR}, vol.~1, no.~2, p.~3, 2022.

\bibitem{c:9}
\BIBentryALTinterwordspacing
A.~Aldoseri, K.~N. Al-Khalifa, and A.~M. Hamouda, ``Re-thinking data strategy and integration for artificial intelligence: Concepts, opportunities, and challenges,'' \emph{Applied Sciences}, vol.~13, no.~12, 2023. [Online]. Available: \url{https://www.mdpi.com/2076-3417/13/12/7082}
\BIBentrySTDinterwordspacing

\bibitem{xu2024fedfa}
H.~Xu, Z.~Zhang, S.~Di, B.~Liu, K.~A. Alharthi, and J.~Cao, ``Fedfa: A fully asynchronous training paradigm for federated learning,'' in \emph{33rd International Joint Conference on Artificial Intelligence, IJCAI 2024}.\hskip 1em plus 0.5em minus 0.4em\relax International Joint Conferences on Artificial Intelligence, 2024, pp. 5281--5288.

\bibitem{zhang2025fedcspc}
Z.~Zhang, S.~Di, K.~Zhao, S.~Jin, D.~Tao, Z.~Ji, B.~Liu, K.~A. Alharthi, J.~Cao, and F.~Cappello, ``Fedcspc: A cross-silo federated learning system with error-bounded lossy parameter compression,'' \emph{IEEE Transactions on Parallel and Distributed Systems}, 2025.

\bibitem{zhang2025fedefsz}
Z.~Zhang, S.~Di, B.~Liu, Z.~Ji, G.~Li, X.~Lu, A.~C. Zhou, K.~A. Alharthi, and J.~Cao, ``Fedefsz: Fair cross-silo federated learning system with error-bounded lossy compression,'' \emph{IEEE Transactions on Parallel and Distributed Systems}, 2025.

\bibitem{c:3}
\BIBentryALTinterwordspacing
R.~Shokri and V.~Shmatikov, ``Privacy-preserving deep learning,'' in \emph{Proceedings of the 22nd ACM SIGSAC Conference on Computer and Communications Security}, ser. CCS '15.\hskip 1em plus 0.5em minus 0.4em\relax New York, NY, USA: Association for Computing Machinery, 2015, p. 1310–1321. [Online]. Available: \url{https://doi.org/10.1145/2810103.2813687}
\BIBentrySTDinterwordspacing

\bibitem{zhang2022mipd}
Z.~Zhang and C.~Wang, ``Mipd: An adaptive gradient sparsification framework for distributed dnns training,'' \emph{IEEE Transactions on Parallel and Distributed Systems}, vol.~33, no.~11, pp. 3053--3066, 2022.

\bibitem{zhang2021sapus}
------, ``Sapus: Self-adaptive parameter update strategy for dnn training on multi-gpu clusters,'' \emph{IEEE Transactions on Parallel and Distributed Systems}, vol.~33, no.~7, pp. 1569--1580, 2021.

\bibitem{zhaorui2022momentum}
Z.~Zhaorui, J.~Zhuoran, and W.~Choli, ``Momentum-driven adaptive synchronization model for distributed dnn training on hpc clusters [j],'' \emph{Journal of Parallel and Distributed Computing}, vol. 159, 2022.

\bibitem{vepakomma2018splitlearninghealthdistributed}
\BIBentryALTinterwordspacing
P.~Vepakomma, O.~Gupta, T.~Swedish, and R.~Raskar, ``Split learning for health: Distributed deep learning without sharing raw patient data,'' 2018. [Online]. Available: \url{https://arxiv.org/abs/1812.00564}
\BIBentrySTDinterwordspacing

\bibitem{c:6}
\BIBentryALTinterwordspacing
------, ``Split learning for health: Distributed deep learning without sharing raw patient data,'' 2018. [Online]. Available: \url{https://arxiv.org/abs/1812.00564}
\BIBentrySTDinterwordspacing

\bibitem{c:7}
\BIBentryALTinterwordspacing
C.~Thapa, M.~A.~P. Chamikara, S.~Camtepe, and L.~Sun, ``Splitfed: When federated learning meets split learning,'' 2022. [Online]. Available: \url{https://arxiv.org/abs/2004.12088}
\BIBentrySTDinterwordspacing

\bibitem{li2019convergence}
X.~Li, K.~Huang, W.~Yang, S.~Wang, and Z.~Zhang, ``On the convergence of fedavg on non-iid data,'' \emph{arXiv:1907.02189}, 2019.

\bibitem{mcmahan2023communicationefficientlearningdeepnetworks}
\BIBentryALTinterwordspacing
H.~B. McMahan, E.~Moore, D.~Ramage, S.~Hampson, and B.~A. y~Arcas, ``Communication-efficient learning of deep networks from decentralized data,'' 2023. [Online]. Available: \url{https://arxiv.org/abs/1602.05629}
\BIBentrySTDinterwordspacing

\bibitem{zhang2024theoreticalanalysisprivacyleakage}
\BIBentryALTinterwordspacing
X.~Zhang and W.~Chen, ``Theoretical analysis of privacy leakage in trustworthy federated learning: A perspective from linear algebra and optimization theory,'' 2024. [Online]. Available: \url{https://arxiv.org/abs/2407.16735}
\BIBentrySTDinterwordspacing

\bibitem{10.1007/978-981-99-9785-5_18}
P.~Lu, X.~Meng, and X.~Liu, ``Fedcmk: An efficient privacy-preserving federated learning framework,'' in \emph{Artificial Intelligence Security and Privacy}, J.~Vaidya, M.~Gabbouj, and J.~Li, Eds.\hskip 1em plus 0.5em minus 0.4em\relax Singapore: Springer Nature Singapore, 2024, pp. 253--271.

\bibitem{rafi2023fairnessprivacypreservingfederatedlearning}
\BIBentryALTinterwordspacing
T.~H. Rafi, F.~A. Noor, T.~Hussain, and D.-K. Chae, ``Fairness and privacy-preserving in federated learning: A survey,'' 2023. [Online]. Available: \url{https://arxiv.org/abs/2306.08402}
\BIBentrySTDinterwordspacing

\bibitem{R1}
C.~Thapa, P.~C.~M. Arachchige, S.~Camtepe, and L.~Sun, ``Splitfed: When federated learning meets split learning,'' in \emph{Proceedings of the AAAI conference on artificial intelligence}, vol.~36, no.~8, 2022, pp. 8485--8493.

\bibitem{R2}
\BIBentryALTinterwordspacing
Y.~Mu and C.~Shen, ``Communication and storage efficient federated split learning,'' 2023. [Online]. Available: \url{https://arxiv.org/abs/2302.05599}
\BIBentrySTDinterwordspacing

\bibitem{R3}
L.~Luo and X.~Zhang, ``Federated split learning via mutual knowledge distillation,'' \emph{IEEE Transactions on Network Science and Engineering}, vol.~11, no.~3, pp. 2729--2741, 2024.

\bibitem{di2025survey}
S.~Di, J.~Liu, K.~Zhao, X.~Liang, R.~Underwood, Z.~Zhang, M.~Shah, Y.~Huang, J.~Huang, X.~Yu \emph{et~al.}, ``A survey on error-bounded lossy compression for scientific datasets,'' \emph{ACM computing surveys}, vol.~57, no.~11, pp. 1--38, 2025.

\bibitem{chenfeddes}
W.~Chen, D.~Zhang, Z.~Chen, Z.~Zhang, G.~Li, S.~Di, and X.~Lu, ``Feddes: A discrete-event simulator for large-scale federated learning.''

\bibitem{liu2025ocelot}
Y.~Liu, S.~Di, J.~Huang, Z.~Zhang, K.~Chard, and I.~Foster, ``Ocelot: An interactive, efficient distributed compression-as-a-service platform with optimized data compression techniques,'' \emph{IEEE Transactions on Parallel and Distributed Systems}, 2025.

\bibitem{huang2024optimized}
J.~Huang, S.~Di, X.~Yu, Y.~Zhai, Z.~Zhang, J.~Liu, X.~Lu, K.~Raffenetti, H.~Zhou, K.~Zhao \emph{et~al.}, ``An optimized error-controlled mpi collective framework integrated with lossy compression,'' in \emph{2024 IEEE International Parallel and Distributed Processing Symposium (IPDPS)}.\hskip 1em plus 0.5em minus 0.4em\relax IEEE, 2024, pp. 752--764.

\bibitem{huang2023c}
J.~Huang, S.~Di, X.~Yu, Y.~Zhai, J.~Liu, K.~Raffenetti, H.~Zhou, K.~Zhao, Z.~Chen, F.~Cappello \emph{et~al.}, ``C-coll: Introducing error-bounded lossy compression into mpi collectives,'' \emph{arXiv preprint arXiv:2304.03890}, 2023.

\bibitem{huang2025zccl}
J.~Huang, S.~Di, X.~Yu, Y.~Zhai, Z.~Zhang, J.~Liu, X.~Lu, K.~Raffenetti, H.~Zhou, K.~Zhao \emph{et~al.}, ``Zccl: Significantly improving collective communication with error-bounded lossy compression,'' \emph{arXiv preprint arXiv:2502.18554}, 2025.

\bibitem{ye2025efficient}
Z.~Ye, S.~Di, J.~Wang, Z.~Zhong, Z.~Zhang, and X.~Yu, ``An efficient gradient-aware error-bounded lossy compressor for federated learning,'' \emph{arXiv preprint arXiv:2511.05770}, 2025.

\bibitem{ngcan}
D.~Ng, D.~Zhang, S.~Di, Z.~Zhang, and X.~Lu, ``Can lossy compression benefit nvme-based io?''

\bibitem{wei2025dynamic}
G.~Wei, Z.~Zhang, J.~Xu, C.~J. Zhang, X.~Yao, and B.~Liu, ``A dynamic virtual memory management system for llms on ai chips,'' in \emph{2025 IEEE 43rd International Conference on Computer Design (ICCD)}.\hskip 1em plus 0.5em minus 0.4em\relax IEEE, 2025, pp. 389--392.

\bibitem{hu2025fastrei}
Z.~Hu, J.~Wang, Z.~Zhong, W.~Zheng, H.~Sharma, J.-S. Park, P.~Kenesei, A.~Miceli, Z.~Zhang, R.~Kettimuthu \emph{et~al.}, ``Fastrei: Fast rare event identification on x-ray data with cross-stage optimizations,'' in \emph{2025 IEEE International Conference on Big Data (BigData)}.\hskip 1em plus 0.5em minus 0.4em\relax IEEE, 2025, pp. 2169--2176.

\bibitem{xu2026deepebc}
J.~Xu, Z.~Zhang, G.~Wei, S.~Di, B.~Liu, X.~Yu, and X.~Lu, ``Deepebc: Compressing the pre-trained llms with error-bounded lossy compression,'' in \emph{Proceedings of the Supercomputing Asia and International Conference on High Performance Computing in Asia Pacific Region Workshops}, 2026, pp. 274--283.

\bibitem{hu2021loralowrankadaptationlarge}
\BIBentryALTinterwordspacing
E.~J. Hu, Y.~Shen, P.~Wallis, Z.~Allen-Zhu, Y.~Li, S.~Wang, L.~Wang, and W.~Chen, ``Lora: Low-rank adaptation of large language models,'' 2021. [Online]. Available: \url{https://arxiv.org/abs/2106.09685}
\BIBentrySTDinterwordspacing

\bibitem{R4}
\BIBentryALTinterwordspacing
Z.~Wang, J.~Liang, R.~He, Z.~Wang, and T.~Tan, ``Lora-pro: Are low-rank adapters properly optimized?'' 2025. [Online]. Available: \url{https://arxiv.org/abs/2407.18242}
\BIBentrySTDinterwordspacing

\bibitem{R5}
\BIBentryALTinterwordspacing
S.-Y. Liu, C.-Y. Wang, H.~Yin, P.~Molchanov, Y.-C.~F. Wang, K.-T. Cheng, and M.-H. Chen, ``Dora: Weight-decomposed low-rank adaptation,'' 2024. [Online]. Available: \url{https://arxiv.org/abs/2402.09353}
\BIBentrySTDinterwordspacing

\bibitem{R6}
\BIBentryALTinterwordspacing
D.~Kalajdzievski, ``A rank stabilization scaling factor for fine-tuning with lora,'' 2023. [Online]. Available: \url{https://arxiv.org/abs/2312.03732}
\BIBentrySTDinterwordspacing

\bibitem{R7}
S.~Hayou, N.~Ghosh, and B.~Yu, ``Lora+: Efficient low rank adaptation of large models,'' 2024.

\bibitem{R8}
\BIBentryALTinterwordspacing
X.~L. Li and P.~Liang, ``Prefix-tuning: Optimizing continuous prompts for generation,'' in \emph{Proceedings of the 59th Annual Meeting of the Association for Computational Linguistics and the 11th International Joint Conference on Natural Language Processing (Volume 1: Long Papers)}, C.~Zong, F.~Xia, W.~Li, and R.~Navigli, Eds.\hskip 1em plus 0.5em minus 0.4em\relax Online: Association for Computational Linguistics, Aug. 2021, pp. 4582--4597. [Online]. Available: \url{https://aclanthology.org/2021.acl-long.353/}
\BIBentrySTDinterwordspacing

\bibitem{R9}
\BIBentryALTinterwordspacing
N.~Houlsby, A.~Giurgiu, S.~Jastrzebski, B.~Morrone, Q.~de~Laroussilhe, A.~Gesmundo, M.~Attariyan, and S.~Gelly, ``Parameter-efficient transfer learning for nlp,'' 2019. [Online]. Available: \url{https://arxiv.org/abs/1902.00751}
\BIBentrySTDinterwordspacing

\end{thebibliography}

\appendix

\section{\textit{SplitFT} Fine-Tuning Workflow}
\label{workflow_appendix}
The detailed workflow of our proposed \textit{SplitFT} is shown in the above Fig. \ref{fig1}. In a local training round, the forward propagation (FP) and backward propagation (BP) processes between the client server and the main server are systematically divided into five distinct steps to ensure efficient communication and computation, optimizing the distributed learning process.

\textbf{(f1) Client-side Forward Propagation:}  
During the local training round \( r \), the client in the randomly generated client list \( N_{\text{client}} \) executes the client-side forward propagation process sequentially. For each client \( i \), a mini-batch of data \( (\mathbf{x}_{i,j}^r, y_{i,j}^r) \) is sampled from the local dataset \( D_{c,i}^{r} \). This data is fed into the client-side pretrained model, parameterized by: \(O_{c,i}^{r} = \left\{ \mathbf{A}_{c,i}^{r,1}, \mathbf{B}_{c,i}^{r,1}, \dots, \mathbf{A}_{c,i}^{r,l_{c,i}}, \mathbf{B}_{c,i}^{r,l_{c,i}} \right\}\), where \( O_{c,i}^{r} \) represents the set of LoRA adapters specific to the client-side pretrained model for client \( i \) at the \( r \)-th training round. The output of the client model, referred to as smashed data, is generated at the cut layer and is denoted as the following formula (\ref{smashed_data}).

\begin{equation}
    \mathbf{\varphi}_i^r = \pi \left( \mathbf{x}_i^r, \mathbf{W}_c, \mathbf{O}_{c,i}^{r-1} \right)
    \label{smashed_data}
\end{equation}

Where \( \pi(x, W, O) \) denotes the computational process and mapping relationship in client-side forward propagation, given the input data \( x \), the pretrained model's weight matrices \( W \), and the trainable LoRA adapters \( O \). The above formula can also be reformulated as the following formula (\ref{smashed_data2}), which describes the output before the client-side cut layer. It can be observed that the memory usage of \( \mathbf{\varphi}_{i,bc}^{r} \) and \( \mathbf{\varphi}_i^r \) on the local client machine remains consistent, irrespective of the rank or configuration of the cut layer and other associated components.

\begin{equation}
    \mathbf{\varphi}_{i,bc}^{r} = \pi \left( \mathbf{x}_i^r, \mathbf{W}_c, \mathbf{O}_{c,i}^{r-1} - \left\{ \mathbf{A}_{c,i}^{r,l_{c,i}}, \mathbf{B}_{c,i}^{r,l_{c,i}} \right\} \right)
    \label{smashed_data2}
\end{equation}

\textbf{(f2) Smashed Data Transmission:}  
The generated smashed data \( \mathbf{\varphi}_i^r \) is transmitted to the main server for further processing in the fine-tuning forward propagation training.


\textbf{(f3) Server-side Forward Propagation and Backward Propagation:}  
Upon receiving the smashed data, the server-side model processes the data, which is parameterized by the formula (\ref{server_para}).

\begin{equation}
O_{s}^{r} = \left\{ \mathbf{A}_{s}^{r,1}, \mathbf{B}_{s}^{r,1}, \mathbf{A}_{s}^{r,2}, \mathbf{B}_{s}^{r,2}, \dots, \mathbf{A}_{s}^{r,M-m}, \mathbf{B}_{s}^{r,M-m} \right\}
    \label{server_para}
\end{equation}

The concatenated input smashed data and the generated prediction labels/tokens are denoted as \( \mathbf{\varphi}^r \) and \( y_{pre}^r \), respectively. In our experiments, we use the Cross Entropy Loss as the loss function, which can be expressed as the following formula (\ref{loss_fun}).

\begin{equation}
    y_{pre}^r = \pi \left( \mathbf{\varphi}_i^r, \mathbf{W}_c, \mathbf{O}_{s}^{r} \right)
    \label{loss_fun}
\end{equation}

The main server computes the loss \( L \) based on \( y_{pre}^r \) and the ground truth labels. It calculates the gradients of the server-side for LoRA adapter decomposition matrices as formula (\ref{grad1}).

\begin{equation}
    \mathbf{A}_s^{r,m} \leftarrow \mathbf{A}_s^{r-1,m} - \gamma_s \mathbf{g}_{A,s}^{r,m}; ~~~ \mathbf{B}_s^{r,m} \leftarrow \mathbf{B}_s^{r-1,m} - \gamma_s \mathbf{g}_{B,s}^{r,m}
    \label{grad1}
\end{equation}


Where \( \gamma_s \) is the server-side learning rate, and \( \mathbf{g}_{A,s}^{r,m} \) and \( \mathbf{g}_{B,s}^{r,m} \) are the gradients of the decomposition matrices \( \mathbf{A}_s^{r,m} \) and \( \mathbf{B}_s^{r,m} \) of the \( m \)-th LoRA adapter on the server-side pretrained model.

\textbf{(f4) Smashed Data Gradients Transmission:}  
After updating the server-side LoRA adapters, the server computes the gradient of the smashed data \( \mathbf{g}_{\phi^r} \) and transmits this gradient back to the corresponding client server to facilitate backward propagation.

\textbf{(f5) Client-side Backward Propagation:}  
Upon receiving gradient \( \mathbf{g}_{\phi^r} \), client performs backward propagation using this gradient to calculate the updates for its local LoRA adapters as formula (\ref{para1}).

\begin{equation}
    \mathbf{g}_{A,c,i}^{r,m} = \mathbf{X}_{c,i}^\top \left( \mathbf{g}_{\phi^r} \mathbf{B}_{c,i}^{r,m} \right); ~~~ \mathbf{g}_{B,c,i}^{r,m} = \left( \mathbf{X}_{c,i} \mathbf{A}_{c,i}^{r,m} \right)^\top \mathbf{g}_{\phi^r}
    \label{para1}
\end{equation}


Where \( \mathbf{X}_{c,i} \) is the local client-side input data, and \( \mathbf{g}_{\phi^r} \) is the gradient of the smashed data transmitted from the server. The gradients for the client-side LoRA adapters \( \mathbf{A}_{c,i}^{r,m} \) and \( \mathbf{B}_{c,i}^{r,m} \) are then used to update these adapters as the formula (\ref{update1}).

\begin{equation}
    \mathbf{A}_{c,i}^{r,m} \leftarrow \mathbf{A}_{c,i}^{r-1,m} - \gamma_c \mathbf{g}_{A,c,i}^{r,m}; ~~~ \mathbf{B}_{c,i}^{r,m} \leftarrow \mathbf{B}_{c,i}^{r-1,m} - \gamma_c \mathbf{g}_{B,c,i}^{r,m}
    \label{update1}
\end{equation}


Where \( \gamma_c \) is the client-side learning rate. The effective weight matrix for the client-side model, incorporating the updated LoRA adapters, is computed as the formula (\ref{update3}).

\begin{equation}
    \mathbf{W}_{\text{client}}^{r} = \mathbf{W}_{\text{base},c} + \mathbf{A}_{c,i}^{r,m} \mathbf{B}_{c,i}^{r,m}
    \label{update3}
\end{equation}

Where \( \mathbf{W}_{\text{base},c} \) is the frozen base weight of the client-side model. This ensures that the client-side model captures task-relevant features while maintaining parameter efficiency.

\subsubsection{Client-side Model Aggregation and Interaction with Server-side Model}
This subsection covers the aggregation of client-side LoRA adapter updates and their interaction with the server-side model, comprising five key steps.

\textbf{(b1) Client-side LoRA Adapters' Update Transmission:}  
After interacting with the main server, each client server \( i \) calculates the changes in its LoRA adapters \( \Delta \mathbf{A}_{c,i}^{R,n} \) and \( \Delta \mathbf{B}_{c,i}^{R,n} \). These updates are then transmitted to the local FedAvg server for aggregation.

\textbf{(b2) Client-side LoRA Adapters' Update Aggregation:}  
Upon receiving updates from all clients, local FedAvg server aggregates the LoRA adapter updates based on the proportion of each client's local training dataset relative to the entire dataset as formula (\ref{update4}).

\begin{equation}
\Delta \mathbf{A}_c^{R,n} = \sum_{i=1}^{N} \frac{\lvert \mathcal{D}_{c,i} \rvert}{\lvert \mathcal{D} \rvert} \Delta \mathbf{A}_{c,i}^{R,n}; ~~~ \Delta \mathbf{B}_c^{R,n} = \sum_{i=1}^{N} \frac{\lvert \mathcal{D}_{c,i} \rvert}{\lvert \mathcal{D} \rvert} \Delta \mathbf{B}_{c,i}^{R,n}
    \label{update4}
\end{equation}


Where \( \Delta \mathbf{A}_c^{R,n} \) and \( \Delta \mathbf{B}_c^{R,n} \) represent the aggregated updates of decomposition matrices \( \mathbf{A} \) and \( \mathbf{B} \) for \( n \)-th client-side LoRA adapter.

\textbf{(b3) Aggregated LoRA Adapters' Update on Client Server:}  
The aggregated updates \( \Delta \mathbf{A}_c^{R,n} \) and \( \Delta \mathbf{B}_c^{R,n} \) are sent back to each client server in \( N_{\text{client}} \), allowing clients to update their respective LoRA adapters accordingly.

\textbf{(b4) Aggregated LoRA Adapters' Update on Main Server:}  
The aggregated updates are transmitted to the main server's base model. The main server evaluates the perplexity and accuracy of the updated base model on each client's training data. Based on these evaluations, the server adjusts the number of layers assigned to each client server to optimize training load and model performance.

\end{document}